\documentclass[]{jfm}

\usepackage{graphicx}
\usepackage{newtxtext}
\usepackage{newtxmath}
\usepackage{natbib}
\usepackage{hyperref}
\hypersetup{
    colorlinks = true,
    urlcolor   = blue,
    citecolor  = black,
}

\newcommand{\RomanNumeralCaps}[1]

\title{Revisiting the hydrodynamic modulation of short surface waves by longer waves}

\author{
  Milan Curcic\aff{1,2}
  \corresp{\email{mcurcic@miami.edu}}
}

\affiliation{
  \aff{1}Rosenstiel School of Marine, Atmospheric, and Earth Science, University of Miami, Miami, FL
  \aff{2}Frost Institute for Data Science and Computing, University of Miami, Coral Gables, FL
}

\begin{document}
\maketitle

\begin{abstract}
Hydrodynamic modulation of short ocean surface waves by longer ambient waves
significantly influences remote sensing, interpretation of \textit{in situ} wave
measurements, and numerical wave forecasting.
This paper revisits the wave crest and action conservation laws and derives
steady, nonlinear, analytical solutions for the change of short-wave wavenumber,
action, and gravitational acceleration due to the presence of longer waves.
We validate the analytical solutions with numerical solutions of the full 
crest and action conservation equations.
The nonlinear analytical solutions of short-wave wavenumber, amplitude, and
steepness modulation significantly deviate from the linear analytical solutions
of \citet{longuet1960changes}, and are similar to the nonlinear numerical
solutions by \citet{longuet1987propagation} and \citet{zhang1990evolution}.
The short-wave steepness modulation is attributed 5/8 to
wavenumber, 1/4 due to wave action, and 1/8 due to effective gravity.
Examining the homogeneity and stationarity requirements for the conservation of
wave action reveals that stationarity is a stronger requirement and is
generally not satisfied for very steep long waves.
We examine the results of \citet{peureux2021unsteady} who found through
numerical simulations that the short-wave modulation grows unsteadily with
each long-wave passage.
We show that this unsteady growth only occurs for homogeneous initial
conditions as a special case and not generally.
The proposed steady solutions are a good approximation of the nonlinear
crest-action conservation solutions in long-wave steepness $\lesssim 0.2$.
Except for a subset of initial conditions, the solutions to the non-linearised
crest-action conservation equations are mostly steady in the reference frame of
the long waves.
\end{abstract}

\begin{keywords}
\end{keywords}


\section{Introduction}

\begin{figure}
\centering
\includegraphics[width=\textwidth]{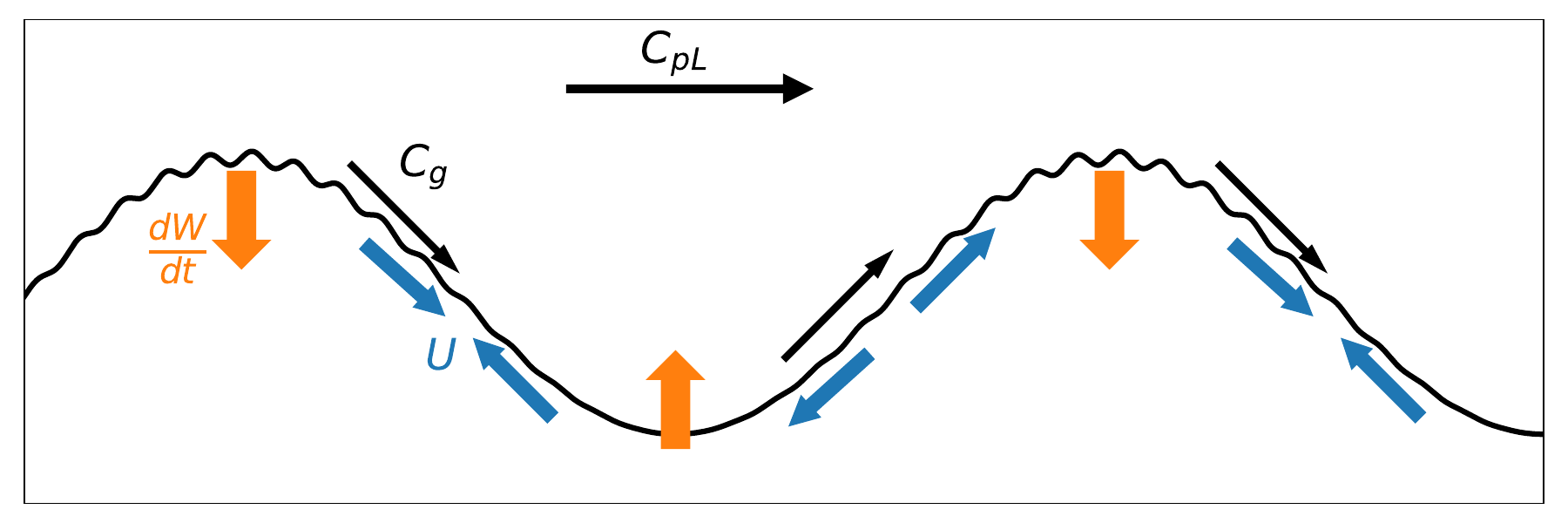}
\caption{
  A diagram of short waves riding on longer waves that propagate with their
  phase-speed $C_{pL}$ from left to right.
  The short waves propagate on the long-wave surface, their action
  moving with the group velocity $C_g$.
  Long waves induce orbital velocities $U$ that cause near-surface convergence
  and divergence on their front and rear faces, respectively.
  They also induce downward and upward centripetal accelerations
  ($dW/dt$) in the crests and troughs, respectively, that
  modulate the effective gravity at the surface.
  As a consequence of surface convergence and divergence and the effective-gravity
  modulation, the short waves become shorter and higher preferentially
  on the long-wave crests.
  The diagram is not to scale.
}
\label{fig:hydrodynamic_modulation_diagram}
\end{figure}

Short ocean surface waves are hydrodynamically modulated by longer swell.
As long waves propagate through a field of shorter waves, their orbital
velocities cause near-surface convergence and divergence on the front and rear
faces of the long waves, respectively.
Following \citet{unna1941white,unna1942waves,unna1947sea},
\citet{longuet1960changes} introduced a steady, approximate solution for
hydrodynamic modulation of short waves by longer waves:

\begin{equation}
\label{eq:lhs1960wavenumber}
\widetilde{k} = k (1 + \varepsilon_L \cos{\psi})
\end{equation}
where $\widetilde{k}$ and $k$ are the modulated and unmodulated short-wave wavenumbers,
respectively, $\varepsilon_L = a_L k_L$ is the steepness of the long wave with
wavenumber $k_L$ and amplitude $a_L$, and $\psi$ is the long-wave phase.
This process is illustrated in Fig. \ref{fig:hydrodynamic_modulation_diagram}.
Hydrodynamic modulation is a key process for the development and
interpretation of remote sensing of ocean surface waves and currents
\citep{keller1975microwave,alpers1978two,hara1994hydrodynamic} and is distinguished
from aerodynamic modulation \citep{donelan1987effect,belcher1999wave,chen2000effects}.
Experimental studies of the general short-wave modulation problem included both
the laboratory \citep[e.g.,][]{keller1975microwave,donelan2010modulation} and field
\citep[e.g.,][]{hara2003observation,plant1977studies} measurements.
\citet{phillips1981dispersion} and \citet{longuet1987propagation} revisited the
hydrodynamic modulation problem by considering nonlinear long waves and
variations in the effective gravity of short waves.
Their results yielded significantly stronger modulation than previously predicted
by \citet{longuet1960changes}.
\citet{henyey1988energy} derived an analytical solution for the modulation of
short waves using Hamiltonian mechanics and reported similar modulation magnitudes
to that of \citet{longuet1987propagation}.
\citet{zhang1990evolution} considered weakly nonlinear short waves
using a nonlinear Schr\" odinger equation and found stronger wavenumber
modulation but weaker amplitude modulation than those predicted by
\citet{longuet1987propagation}.
Aside from the early linear solutions by \citet{longuet1960changes}, the variety
of analytical and numerical frameworks (Eulerian crest and action conservation,
Hamiltonian, Schr\" odinger) that tackle the same fundamental
physics and that relatively closely reproduce the steady solutions for short-wave
modulation suggests that the solutions are robust.
Nevertheless, alternative analytical and numerical approaches to the modulation
problem remain of interest, especially if higher degrees of nonlinearity can be
considered using simpler approaches.

All modulation solutions mentioned thus far are steady in the reference frame
of the long wave---short waves become shorter and higher (and thus, steeper)
preferentially on the long-wave crests.
Conversely, they elongate and flatten preferentially in the long-wave troughs.
Recently, \citet{peureux2021unsteady} asked whether the steady
(\textit{i.e.}, stationary; not depending on time) solutions
of short-wave modulation are appropriate by simulating the full wave conservation
equations.
They found that the short waves grow unsteadily due to the propagation of
longer waves, suggesting that the steady-state solutions may only be valid for
a few long-wave periods before the short waves begin to break due to excessive
steepness.
However, the unsteadiness of their solutions occurs in a specific scenario in
which a long-wave train suddenly appears to modulate a uniform field of short
waves.
The numerical simulations stabilise if the short-wave field is initialised
from existing steady solutions, like those by \citet{longuet1960changes}.
Nevertheless, their results do put into question whether the short-wave
modulation predicted by theory is generally steady over many long-wave periods
or not.
Looking at the conservation equations alone, it is not obvious that it is.

\begin{table}
\begin{center}
\def~{\hphantom{0}}
\begin{tabular}{cccccc}
Reference   & Analysis   & Solution & Solution    & Nonlinear  & Nonlinear   \\
            & framework  & type     & steadiness  & long waves & short waves \\
\hline
\citet{longuet1960changes} & Velocity potential  & Analytical & Steady & No  & No  \\
\hline
\citet{phillips1981dispersion} & Crest and action & Numerical & Steady & Yes & No  \\
                               & conservation     &           &        &     &     \\
\hline
\citet{longuet1987propagation} & Crest and action & Numerical & Steady & Yes & No  \\
                               & conservation     &           &        &     &     \\
\hline
\citet{henyey1988energy} & Hamiltonian & Analytical & Steady & Yes & No  \\
\hline
\citet{zhang1990evolution} & Schr\" odinger & Numerical & Steady & Yes & Yes \\
\hline
\citet{peureux2021unsteady} & Crest and action & Numerical & Unsteady & No  & No  \\
                            & conservation     &           &          &     &     \\
\hline
This paper  & Crest and action & Analytical    & Steady       & Yes & No  \\
            & conservation     & and numerical & and unsteady &     &     \\
\hline
\end{tabular}
\caption{
  Non-exhaustive summary of prior and present (this paper) approaches to
  calculating the modulation of short waves by long waves.
  The solution steadiness refers to whether the solution depends on time
  (unsteady) or not (steady).
}
\label{table:modulation_literature_summary}
\end{center}
\end{table}

In this paper, we revisit this problem from the perspective of wave action
conservation but consider the long-wave slope to be significant enough to
require evaluating the short-waves at the long-wave surface rather than at the
mean water level.
This allows us to derive alternative, steady, nonlinear solutions for the
modulation of short waves, within the limits of small $\varepsilon_L$.
The analytical solutions yield modulations that are stronger than the original
solution by \citet{longuet1960changes}, but similar to the nonlinear numerical
solutions by \citet{longuet1987propagation} and \citet{zhang1990evolution} for
$\varepsilon_L \lesssim 0.2$.
To validate the steadiness of the analytical solutions, we perform numerical
simulations of the full wave crest and action conservation laws, and compare
the results to the analytical solutions presented here, as well as the prior
numerical results.
The numerical simulations are used to determine the validity of the approximate
steady solution and to investigate the unsteadiness of the short-wave modulation
found by \citet{peureux2021unsteady}.
The main prior studies on hydrodynamic modulation, and their assumptions and
approaches to the solution, are summarised in Table
\ref{table:modulation_literature_summary}.
We first present the governing equations in \S\ref{section:governing_equations},
and then the analytical solutions for the modulation of short waves by long waves
in \S\ref{section:analytical_solutions}.
Then, we describe the numerical simulations and their results in
\S\ref{section:numerical_solutions}.
Finally, we discuss the results and their implications in
\S\ref{section:discussion} and conclude the paper in \S\ref{section:conclusions}.

\section{Governing equations}
\label{section:governing_equations}

The flow is inviscid, irrotational, and incompressible, and without sources or
sinks.
In these conditions, linear, deep-water, surface gravity waves obey the
dispersion relation:

\begin{equation}
\label{eq:dispersion}
\omega = \sqrt{gk} + k U
\end{equation}
where $\omega$ is the angular frequency in a fixed reference frame, $g$ is the
gravitational acceleration, $k$ is the wavenumber, and $U$ is the mean advective
current in the direction of the wave propagation.
A current in the direction of the waves increases their absolute frequency
without changing their wavenumber.
An important limitation here is that $U$ must be slowly varying on the scales of
the wave period \citep{bretherton1968wavetrains}.
From the perspective of short waves riding on long waves, the advective current
is the horizontal near-surface orbital velocity of the long wave.
Strictly, shear must be considered when determining the advective velocity for
a wave of any given wavenumber \citep{stewart1974hf,ellingsen2014ship}.
Here, we assume the surface velocity to be the advective one for short waves,
and demonstrate later in $\S$\ref{sec:wave_action_conservation} that this is
valid for sufficient scale separation between the short and long waves.
The evolution of the short-wave wavenumber is described by the conservation of
wave crests \citep[e.g.,][]{whitham1974dispersive,phillips1981dispersion}:

\begin{equation}
\label{eq:wave_crests}
\dfrac{\partial k}{\partial t}
+ \dfrac{\partial \omega}{\partial x}
= 0
\end{equation}
where $t$ and $x$ are time and space, respectively.
This relation states that the wavenumber must change locally when the frequency
varies in space.

(\ref{eq:dispersion}) and (\ref{eq:wave_crests}) describe the evolution of
the short-wave wavenumber.
To describe the evolution of the wave amplitude, use the wave action balance
\citep{bretherton1968wavetrains} in absence of sources (e.g. due to wind) and
sinks (e.g. due to whitecapping):

\begin{equation}
\label{eq:wave_action}
\dfrac{\partial N}{\partial t}
+ \dfrac{\partial}{\partial x} \left[\left(C_g + U\right)N\right]
= 0
\end{equation}
where $N$ is the action of short waves defined as the ratio of their
energy to their intrinsic frequency, $E/\sigma$,
and $C_g = 1/2\sqrt{g/k}$ is their group speed in deep water.
The wave action expressed as $E/\sigma$ is commonly taken as
conservative in wave modulation theory
\citep{phillips1981dispersion,longuet1987propagation}, however as
\citet{dysthe1988orbiting} show, it is conservative only for gently sloped waves.
We describe this and other requirements in more detail further below and show
in what long-wave conditions they are satisfied.
The wave crest conservation (\ref{eq:wave_crests}) follows from the assumption
of a conserved wave phase, and is another prerequisite for wave action
conservation \citep[e.g.,][]{whitham1974dispersive}.
Although the absence of wave growth and dissipation is not generally realistic in
the open ocean, it is a useful approximation here to isolate the effects
of hydrodynamic modulation solely due to the motion of the long waves.
The governing equations (\ref{eq:dispersion})-(\ref{eq:wave_action}) are
derived in Appendix \ref{appendix:derivation}.

Short waves that ride on the surface of the longer waves move in an accelerated
reference frame (due to the centripetal acceleration at the long-wave surface)
and thus experience effective gravitational acceleration that varies in space
and time \citep{phillips1981dispersion,longuet1986eulerian,longuet1987propagation}.
Inserting (\ref{eq:dispersion}) into (\ref{eq:wave_crests}) and noting that:

\begin{equation}
\label{eq:dispersion_derivative}
\dfrac{\partial \sqrt{gk}}{\partial x} =
\dfrac{1}{2} \sqrt{\dfrac{g}{k}} \dfrac{\partial k}{\partial x} + \dfrac{1}{2} \sqrt{\dfrac{k}{g}} \dfrac{\partial g}{\partial x}
\end{equation}
we get:

\begin{equation}
\label{eq:wavenumber}
\dfrac{\partial k}{\partial t}
+ \left(C_g + U\right) \dfrac{\partial k}{\partial x}
+ k \dfrac{\partial U}{\partial x}
+ \dfrac{1}{2} \sqrt{\dfrac{k}{g}} \dfrac{\partial g}{\partial x}
= 0
\end{equation}
From left to right, the terms in this equation represent the change in wavenumber
due to the propagation and advection by ambient current $U$, the convergence of
the ambient current, and the spatial inhomogeneity of the gravitational acceleration.
In a non-accelerated reference frame (i.e. in absence of longer waves), the last
term vanishes.
It is otherwise necessary to conserve the wave crests.

For the wave action, expand (\ref{eq:wave_action}) to get:

\begin{equation}
\label{eq:wave_action2}
\dfrac{\partial N}{\partial t}
+ \left(C_g + U\right) \dfrac{\partial N}{\partial x}
+ N \dfrac{\partial U}{\partial x}
+ N \dfrac{\partial C_g}{\partial x}
= 0
\end{equation}
This equation is similar to (\ref{eq:wavenumber}), except for the last term
which represents the inhomogeneity of the group speed.
As \citet{peureux2021unsteady} explained, the presence of this term causes
unsteady growth of wave action in infinite long-wave trains.
(\ref{eq:wavenumber}) and (\ref{eq:wave_action2}) are the governing equations
that we approximate and solve analytically in \S\ref{section:analytical_solutions},
and numerically in their full form in \S\ref{section:numerical_solutions}.
This system of equations is semi-coupled, meaning that
it is possible to solve (\ref{eq:wavenumber}) on its own, but solving
(\ref{eq:wave_action2}) requires also the solution of (\ref{eq:wavenumber}).
In other words, the wave kinematics (the wavenumber distribution) are not
concerned with the wave action.
In contrast, the wave action is governed, among others, by the
group velocity and thus depends on the wavenumber.
This convenient property is only possible if we consider the linear form
of the dispersion relation, that is, if the series are truncated at the
second-order solution in the Stokes expansion.
For a third-order expansion, an additional term that is proportional to the
square of the wave steepness appears in the dispersion relation.
In that case, equations (\ref{eq:wavenumber}) and (\ref{eq:wave_action2})
become fully coupled and the analytical solution for the wavenumber is no longer
feasible.

For the ambient forcing, we consider a train of monochromatic long waves
defined, to the first order in $\varepsilon_L$, by the elevation
$\eta_L = a_L \cos{\psi}$, where $\psi = k_L x - \sigma_L t$ is the long-wave
phase and $\sigma_L$ its angular frequency.
The velocity potential of the long wave is:

\begin{equation}
\label{eq:phi}
\phi = \dfrac{a_L \omega_L}{k_L} e^{k_L z} \sin{\psi} + \mathcal{O}(\varepsilon_L^4)
\end{equation}
where $z$ is the vertical distance from the mean water level that is positive
upwards.
When evaluated at the mean water level ($z=0$), (\ref{eq:phi}) is accurate to
the third order in $\varepsilon_L$ in deep water because the second- and
third-order terms in the Stokes expansion series are exactly zero.
The long-wave induced horizontal and vertical orbital velocities are:

\begin{equation}
\label{eq:U_L}
U =
\frac{\partial \phi}{\partial x} =
a_L \omega_L e^{k_L z} \cos{\psi} + \mathcal{O}(\varepsilon_L^4)
\end{equation}

\begin{equation}
\label{eq:W_L}
W =
\frac{\partial \phi}{\partial z} =
a_L \omega_L e^{k_L z} \sin{\psi} + \mathcal{O}(\varepsilon_L^4)
\end{equation}
Evaluating these velocities at the wave surface $z = \eta_L$ rather than at
the mean water level $z=0$ is a departure from the linear theory, making the
treatment of the long waves here as weakly nonlinear.
As \citet{zhang1990evolution} point out, this is appropriate to do when the
amplitude of the long waves is of the similar magnitude or larger than the
wavelength of the short waves.
However, evaluating these velocities at the free surface $z = \eta_L$ introduces
an additional error of $\mathcal{O}(\varepsilon_L^2)$.
The surface velocity errors are described in more detail in Appendix
\ref{appendix:surface_velocity_errors}.

(\ref{eq:wavenumber})-(\ref{eq:wave_action2}) are the governing equations used
in this paper, with (\ref{eq:U_L})-(\ref{eq:W_L}) describing the long-wave
induced ambient velocity forcing for the short waves. 
Rather than deriving the modulation solution from the velocity
potential of two waves, as \citet{longuet1960changes} did, starting from the
crest-action conservation balances allows for a simpler derivation.

\section{Analytical solutions}
\label{section:analytical_solutions}

We now describe the analytical solutions for the short-wave wavenumber,
effective gravitational acceleration, amplitude, and steepness in the presence
of long waves.
The wavenumber is derived by linearising the conservation of wave crests
(\ref{eq:wave_crests}) and neglecting the spatial gradients of $k$ and $g$.
The effective gravity is derived by projecting the long-wave induced surface
accelerations onto the local vertical axis (that is, the axis normal to the
long-wave surface), akin to \citet{zhang1990evolution}.
The amplitude is derived by first evaluating the short-wave action from the
action conservation (\ref{eq:wave_action}) and then applying the modulated
effective gravity to obtain the amplitude.
From the modulated wavenumber, amplitude, and effective gravity, the steepness,
frequency, and phase speed readily follow from the usual wave relationships.
Going forward we will use the tilde over the variables to denote the modulated
quantities (\textit{e.g.} $\widetilde{k}$ for wavenumber).

\subsection{Wavenumber modulation}
\label{subsection:wavenumber_modulation}

Solving analytically for the wavenumber requires dropping the spatial derivatives
of $k$ in (\ref{eq:wavenumber}) and assuming homogeneous $g$:

\begin{equation}
\label{eq:wavenumber2}
\dfrac{\partial k}{\partial t}
= - k \dfrac{\partial U}{\partial x}
\end{equation}
Although \citet{peureux2021unsteady} pointed out that the solution by
\citet{longuet1960changes} requires assuming homogeneous group speed of the
short-wave field, in fact, it also requires assuming no horizontal propagation
and advection of the short waves.
Combine (\ref{eq:U_L}) and (\ref{eq:wavenumber2}) and integrate in time to get:

\begin{equation}
\label{eq:k_short_exact}
\frac{\widetilde{k}}{k} = e^{\varepsilon_L \cos{\psi} e^{\varepsilon_L \cos{\psi}}}
\end{equation}

\begin{figure}
\centering
\includegraphics[width=0.8\textwidth]{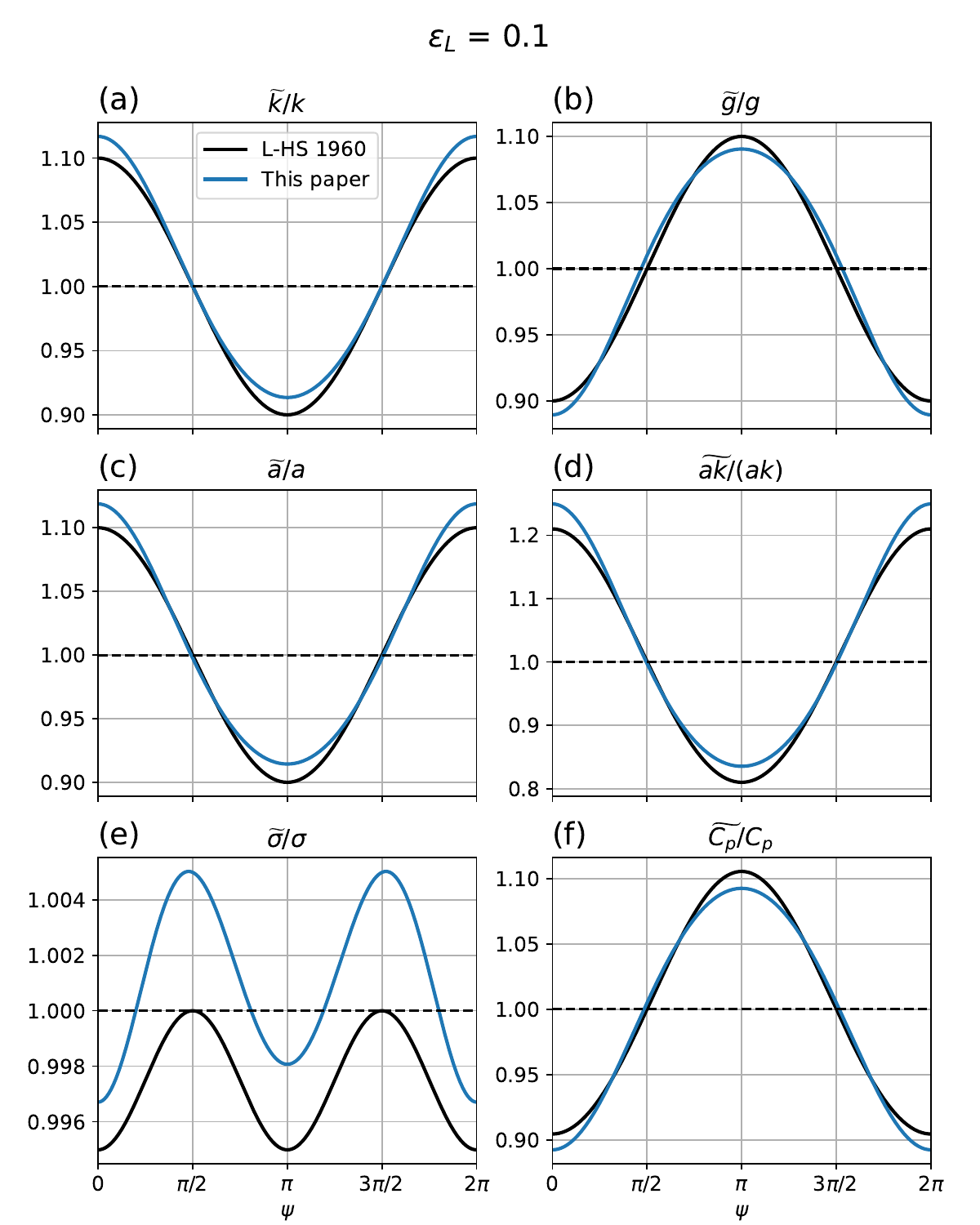}
\caption{
  Steady analytical solutions for the modulation of short-wave (a) wavenumber, (b)
  gravitational acceleration, (c) amplitude, (d) steepness, (e) intrinsic
  frequency, and (f) intrinsic phase speed as function of long-wave phase for
  $\varepsilon_L = 0.1$, based on
  \citet{longuet1960changes} (L-HS 1960, black) and this paper (blue).
  Long-wave crest and trough are located at $\psi = 0$ and $\psi = \pi$,
  respectively.
}
\label{fig:analytical_solutions_ak0.1}
\end{figure}

Notice that the Taylor expansion of (\ref{eq:k_short_exact}) to the first order
recovers the original solution by \citet{longuet1960changes}:

\begin{equation}
\label{eq:k_short_lhs}
\frac{\widetilde{k}}{k} = 1 + \varepsilon_L \cos{\psi}
\end{equation}
Coming from the conservation of crests, as opposed to the velocity potential
superposition, the solution by \citet{longuet1960changes} requires two
approximations:
First, evaluating the long-wave velocity at $z = 0$ rather than $z = \eta_L$; and
second, truncating the Taylor expansion series beyond $\mathcal{O}(\varepsilon_L)$.
These approximations underestimate the short-wave modulation magnitude
(Fig. \ref{fig:analytical_solutions_ak0.1}).
At the wave crest, (\ref{eq:k_short_exact}) yields the wavenumber modulation
that is 16.9\%, 38.5\%, 66.4\%, and 104\% larger than that of
\citet{longuet1960changes} for $\varepsilon_L = 0.1, 0.2, 0.3, 0.4$, respectively.
Although not immediately obvious from the expressions, the modulated wavenumber
(\ref{eq:k_short_lhs}) is conserved across the long-wave phase, whereas
the higher-order solution (\ref{eq:k_short_exact}) is not.
This is because combining the orbital velocities evaluated at $z=\eta_L$ and the
linearised conservation of wave crests in (\ref{eq:wavenumber}) creates excess
short-wave wavenumber.
The solution of (\ref{eq:wavenumber}) must be conservative across the long-wave
phase, as we will see later from the numerical simulations.
The steady analytical solutions for modulation of short-wave effective gravity,
amplitude, steepness, intrinsic frequency, and phase speed are also shown in
Fig. \ref{fig:analytical_solutions_ak0.1}, for reference, however, their
formulations appear in the following subsections.
Although the analytical solutions by \citet{longuet1960changes} have been updated
by the numerical results of \citet{longuet1987propagation}, they are here
nevertheless relevant to use as reference for the proposed analytical solutions.
This comparison allows quantifying solely the effect of evaluating the
long-wave properties at the surface rather than at the mean water level.

\subsection{Gravity modulation}
\label{subsection:gravity_modulation}

Three distinct effects contribute to the modulation of the effective gravity of
short waves in the presence of long waves.
First, long waves induce orbital (centripetal) accelerations on their surface,
so the short waves ride in an accelerated reference frame and experience effective
gravity that is lower at the crests and higher in the troughs
\citep{longuet1986eulerian,longuet1987propagation}.
This is an $\mathcal{O}(\varepsilon_L)$ effect.
Second, it is important to distinguish between the Eulerian and Lagrangian effective
gravities \citep{longuet1986eulerian}.
The Eulerian gravity is that which would be measured by a fixed wave staff,
for example, and features sharp and strong minima at the crests and broad
and weak maxima in the troughs.
In contrast, the Lagrangian gravity is that which is experienced by a short wave
group that propagates with its own group speed and is advected by the ambient,
long-wave induced, velocity.
The Lagrangian velocity is overall smoother and more attenuated compared to
its Eulerian counterpart because the local slope as experienced by the
travelling short-wave group is gentler.
The two gravities are related by:

\begin{equation}
\label{eq:gravity_modulation_lagrangian}
\widetilde{g_l} =
g + \frac{dW}{dt} =
g + \frac{\partial W}{\partial t} + \left(C_g + U\right) \frac{\partial W}{\partial x} =
\widetilde{g_e} + \left(C_g + U\right) \frac{\partial W}{\partial x}
\end{equation}
where $l$ and $e$ subscripts serve to denote "Lagrangian" and "Eulerian",
respectively.
Although \citet{longuet1986eulerian} thoroughly described the differences
between the Eulerian and Lagrangian effective gravities, he did so for
a fully-nonlinear irrotational gravity wave, which can be computed numerically
but not analytically.
Note also that when evaluating the Lagrangian effective gravity,
\citet{longuet1987propagation} considered the orbital motion of the long wave
but not the propagation speed of the short-wave group, which can exceed the
ambient velocity $U$ depending on the properties of the short and long waves
considered.
Group speed of the short waves aside, the Lagrangian correction to the gravity
being an advective term is $\mathcal{O}(\varepsilon_L^2)$.
Third, the projection of the long-wave induced horizontal acceleration in the
curvilinear coordinate system contributes small but non-negligible effects on
the short-wave effective gravity \citep{phillips1981dispersion,zhang1990evolution}.
These three effects are completely described by projecting the gravitational
acceleration vector from the coordinate that is perpendicular to the mean water
surface ($z=0$) to that which is perpendicular to the long-wave surface
($z=\eta_L$), and likewise for the long-wave orbital accelerations:

\begin{equation}
\label{eq:gravity_modulation_general}
\frac{\widetilde{g}}{g}
  = \cos{\alpha} 
  + \frac{1}{g} \dfrac{dW_{z=\eta_L}}{dt} \cos{\alpha}
  + \frac{1}{g} \dfrac{dU_{z=\eta_L}}{dt} \sin{\alpha}
\end{equation}

\begin{equation}
\label{eq:local_slope}
\alpha = \tan^{-1}{\dfrac{\partial \eta}{\partial x}}
\end{equation}
where $\alpha$ is the local slope of the long-wave surface.
For short waves on a linear long wave, the Eulerian gravity modulation is:

\begin{equation}
\label{eq:gravity_modulation_linear}
\frac{\widetilde{g_e}}{g}
  = 
  \frac{
    1 - \varepsilon_L \cos{\psi} e^{\varepsilon_L \cos{\psi}}
    \left[ 1 + \left(\varepsilon_L \sin{\psi}\right)^2 e^{\varepsilon_L \cos{\psi}} \right]
  }
  {\sqrt{\left(\varepsilon_L \sin{\psi}\right)^2 + 1}}
\end{equation}
The same derivation can be carried out for the Lagrangian gravity
modulation following (\ref{eq:gravity_modulation_lagrangian})-(\ref{eq:local_slope}),
and is omitted here for brevity.
Curvilinear effects on effective gravity can be removed altogether by setting
$\alpha = 0$.
In that case, (\ref{eq:gravity_modulation_linear}) simplifies to:

\begin{equation}
\label{eq:gravity_modulation_linear_no_curvature}
\frac{\widetilde{g_e}}{g} = 1 - \varepsilon_L \cos{\psi} e^{\varepsilon_L \cos{\psi}}
\end{equation}
If we evaluate the orbital accelerations at the mean water level $z=0$,
this further simplifies to:

\begin{equation}
\label{eq:gravity_modulation_linear_no_curvature_mean_level}
\frac{\widetilde{g_e}}{g} = 1 - \varepsilon_L \cos{\psi}
\end{equation}
which is the form used by \citet{longuet1960changes} and
\citet{peureux2021unsteady}, for example.
The Lagrangian correction in (\ref{eq:gravity_modulation_lagrangian}) is:

\begin{equation}
\label{eq:gravity_modulation_lagrangian_correction}
U \frac{\partial W}{\partial x} = g \varepsilon_L^2 \cos^2{\psi} e^{2 \varepsilon_L \cos{\psi}}
+ \mathcal{O}(\varepsilon_L^3)
\end{equation}
and so it becomes important only for large $\varepsilon_L$.

The general form of the gravity modulation (\ref{eq:gravity_modulation_general})
is convenient because it allows us to easily attribute the modulation to
different processes.
It also makes it straightforward to evaluate the gravity modulation for
different long-wave forms, by evaluating the orbital accelerations and the local
slope for each wave form.
For a third-order Stokes wave, for example, the elevation is:

\begin{equation}
\label{eq:eta_stokes}
\eta_{St} = a_L \left[
  \cos{\psi} +
  \dfrac{1}{2} \varepsilon_L \cos{2\psi} +
  \varepsilon_L^2 \left( \dfrac{3}{8} \cos{3\psi} - \dfrac{1}{16} \cos{\psi} \right)
\right] + \mathcal{O}(\varepsilon_L^4)
\end{equation}
Without considering the curvilinear effects in (\ref{eq:gravity_modulation_general}),
the gravitational acceleration at the surface of a Stokes wave is, to the third
order:

\begin{equation}
\label{eq:gravity_modulation_stokes}
\frac{\widetilde{g}}{g} =
\left\{
  1 - \varepsilon_L e^{k \eta_{St}}
  \left[ \cos{\psi} -
    \varepsilon_L \sin{\psi} \left(
      \sin{\psi}
      + \varepsilon_L \sin{2\psi}
      - \dfrac{1}{16} \varepsilon_L^2 \sin{\psi}
      + \dfrac{9}{8} \varepsilon_L^2 \sin{3\psi}
    \right)
  \right]
\right\}
\end{equation}
The equivalent expression that includes the curvilinear effects can be derived
by evaluating the velocities and the local slope $\alpha$ at $z = \eta_{St}$,
and inserting them into (\ref{eq:gravity_modulation_general}):

\begin{equation}
\label{eq:gravity_modulation_general_stokes}
\frac{\widetilde{g}}{g}
  = \cos{\alpha_{St}}
  + \dfrac{dW_{z=\eta_{St}}}{dt} \cos{\alpha_{St}}
  + \dfrac{dU_{z=\eta_{St}}}{dt} \sin{\alpha_{St}}
\end{equation}

\begin{equation}
\label{eq:local_slope_stokes}
\alpha_{St} = \tan^{-1}{\dfrac{\partial \eta_{St}}{\partial x}}
\end{equation}

\begin{figure}
\centering
\includegraphics[width=\textwidth]{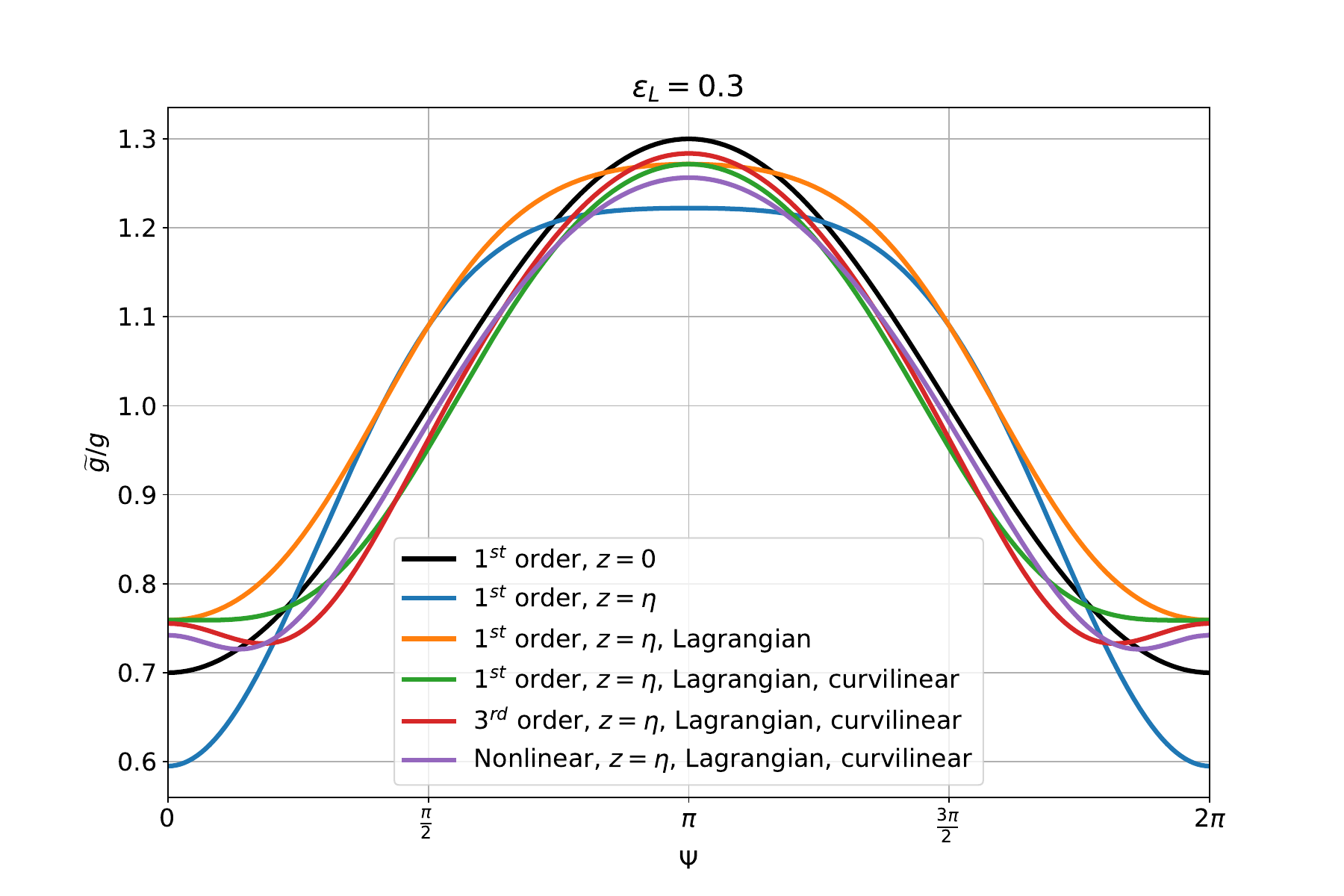}
\caption{
  Analytical solutions for the effective gravitational acceleration
  modulation by long waves, as function of the long-wave phase, for $\varepsilon_L = 0.3$.
  Black is for Eulerian gravity of a linear wave evaluated at $z=0$;
  blue is the same as black but evaluated at $z=\eta$;
  orange is the same as blue but for Lagrangian gravity;
  green is the same as orange but in a curvilinear reference frame;
  red is the same as green but for a third-order Stokes wave;
  and purple is the same as red but for a fully-nonlinear wave.
  The elevation and surface velocities for the fully-nonlinear wave are computed
  following \citet{clamond2018accurate}.
  Long-wave crest and trough are located at $\psi = 0$ and $\psi = \pi$,
  respectively.
}
\label{fig:effective_gravities}
\end{figure}

Let us now examine the contributions of the different terms to the short-wave
effective gravity modulation.
Fig. \ref{fig:effective_gravities}a shows the effective gravity modulation of
short waves as a function of the long-wave phase for $\varepsilon_L = 0.3$, for
a variety of the gravity modulation forms.
Although quite steep, this value of $\varepsilon_L$ allows easier visualisation
of the modulation differences.
To $\mathcal{O}(\varepsilon_L)$, the gravity modulation is out of phase
with the long-wave elevation, with the short waves experiencing a lower effective
gravity on the long-wave crests and a higher effective gravity in the troughs.
Evaluating the orbital accelerations at $z=\eta$ and the Lagrangian gravity
correction each have an $\mathcal{O}(\varepsilon_L^2)$ contribution but of
opposite signs.
The former amplifies the gravity modulation while the latter attenuates it.
The curvilinear effects are significant at the front and rear faces of the long
wave but not near the crests and troughs.
The long-wave surface slope has a $\mathcal{O}(\varepsilon_L^2)$ effect.
Naturally, the slope is exactly zero at the crests and the troughs, and its
effects are largest at the front and rear faces of the long wave, where the
covariance between the local slope and the horizontal accelerations is largest.
Evaluating the velocities and accelerations at the surface of a third-order Stokes
wave causes a correction mostly at the faces of the long wave.
Finally, numerically computing the surface velocities and accelerations of a
fully-nonlinear wave allows us to exclude uncertainties associated with the
truncation errors of the Stokes expansion at first and third orders in
$\varepsilon_L$.
At this steepness, the fully-nonlinear solution is marginally different from
that of the third-order Stokes wave
(see also Appendix \ref{appendix:surface_velocity_errors} for the errors
in surface velocity estimates of the linear and third-order Stokes
approximations).

\begin{figure}
\centering
\includegraphics[width=\textwidth]{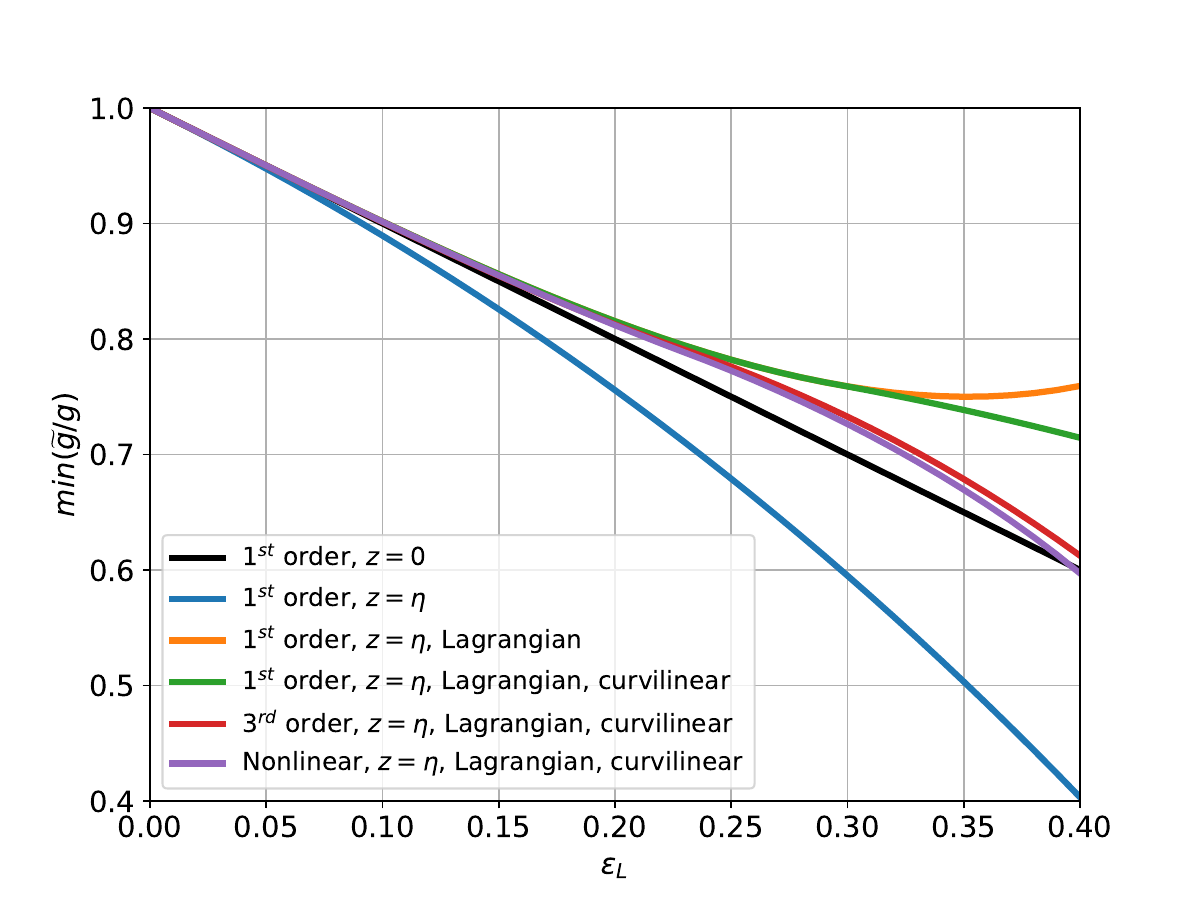}
\caption{
  As Fig. \ref{fig:effective_gravities} but showing the minimum short-wave
  gravity modulation as function of long-wave steepness $\varepsilon_L$.
}
\label{fig:effective_gravities_min}
\end{figure}

It is instructive to also examine the minimum values of the
effective gravity as function of long-wave steepness $\varepsilon_L$
(Fig. \ref{fig:effective_gravities_min}), as the minima occur at or
near the long-wave crests where the short-wave modulation is largest.
Without the Lagrangian correction to the effective gravity, the gravity
modulation is overestimated; for accelerations evaluated at the surface of a
linear wave, the maximum gravity reduction is 60\% at $\varepsilon_L = 0.4$.
The Lagrangian correction attenuates the gravity modulation, making the
gravity reduction at the crests $\approx 25\%$.
As the slope effects on the effective gravity exist only away from crests and
troughs, they may be neglected in the steady solutions where modulation is
typically evaluated at the long-wave crests.
Here, the slope effect causes a difference in the minimum effective gravity
for $\varepsilon_L \gtrsim 0.3$.
The effective gravity reduction at the crests in the case of a fully-nonlinear
wave is $\approx$ 10\%, 19\%, 27\%, and 40\%, 
for $\varepsilon_L$ of 0.1, 0.2, 0.3, and 0.4, respectively.
These are similar to the fully-nonlinear numerical estimates of
\citet{longuet1986eulerian,longuet1987propagation}.

\subsection{Amplitude and steepness modulation}
\label{subsection:amplitude_modulation}

The modulation of short-wave amplitude can be derived in a similar way as we did
for the wavenumber, except that here we linearise the wave action balance
(\ref{eq:wave_action2}) and integrate it in time:

\begin{equation}
\label{eq:wave_action_modulation}
\frac{\widetilde{N}}{N} = e^{\varepsilon_L \cos{\psi} e^{\varepsilon_L \cos{\psi}}}
\end{equation}

Since wave action is energy divided by the intrinsic frequency $\sigma$,
and energy scales with $ga^2$, the modulation of the amplitude is related to the
modulations of gravity, wavenumber, and action:

\begin{equation}
\label{eq:wave_amplitude_modulation}
\dfrac{\widetilde{a}}{a} = \sqrt{
  \dfrac{g}{\widetilde{g}}
  \dfrac{\widetilde{\sigma}}{\sigma}
  \dfrac{\widetilde{N}}{N}}
=
  \left( \dfrac{\widetilde{k}}{k} \right)^{\frac{1}{4}}
  \left( \dfrac{\widetilde{N}}{N} \right)^{\frac{1}{2}}
  \left( \dfrac{\widetilde{g}}{g} \right)^{-\frac{1}{4}}
\end{equation}
To $\mathcal{O}(\varepsilon_L)$, and using (\ref{eq:k_short_exact})
and (\ref{eq:wave_action_modulation}), the above simplifies to the result of
\citet{longuet1960changes}:

\begin{equation}
\label{eq:wave_amplitude_modulation_order1}
\dfrac{\widetilde{a}}{a} = 
  \left( \dfrac{\widetilde{g}}{g} \right)^{-\frac{1}{4}}
  \left( 1 + \varepsilon_L \right)^{\frac{3}{4}}
  \approx \frac{(1 + \varepsilon_L)^{\frac{3}{4}}}{(1 - \varepsilon_L)^{\frac{1}{4}}}
  \approx 1 + \varepsilon_L
\end{equation}

The amplitude and steepness modulation of the short waves as a function of the
long-wave phase for $\varepsilon_L = 0.1$ are compared to the
\citet{longuet1960changes} solutions in Fig.
\ref{fig:analytical_solutions_ak0.1} panels (c) and (d), respectively.
Also shown are the modulation of the short-wave intrinsic frequency and phase
speed in panels (e) and (f), respectively.
As the wavenumber and gravity modulations are both $\mathcal{O}(\varepsilon_L)$
and out of phase, the intrinsic frequency modulation is
$\mathcal{O}(\varepsilon_L^2)$, positive on the long-wave faces and negative
on the crests and troughs.
For the short-wave phase speed modulation, the wavenumber and gravity modulations
work in tandem and cause the short-wave phase speed to be reduced on the long-wave
crests and increased in the troughs.

\begin{figure}
\centering
\includegraphics[width=\textwidth]{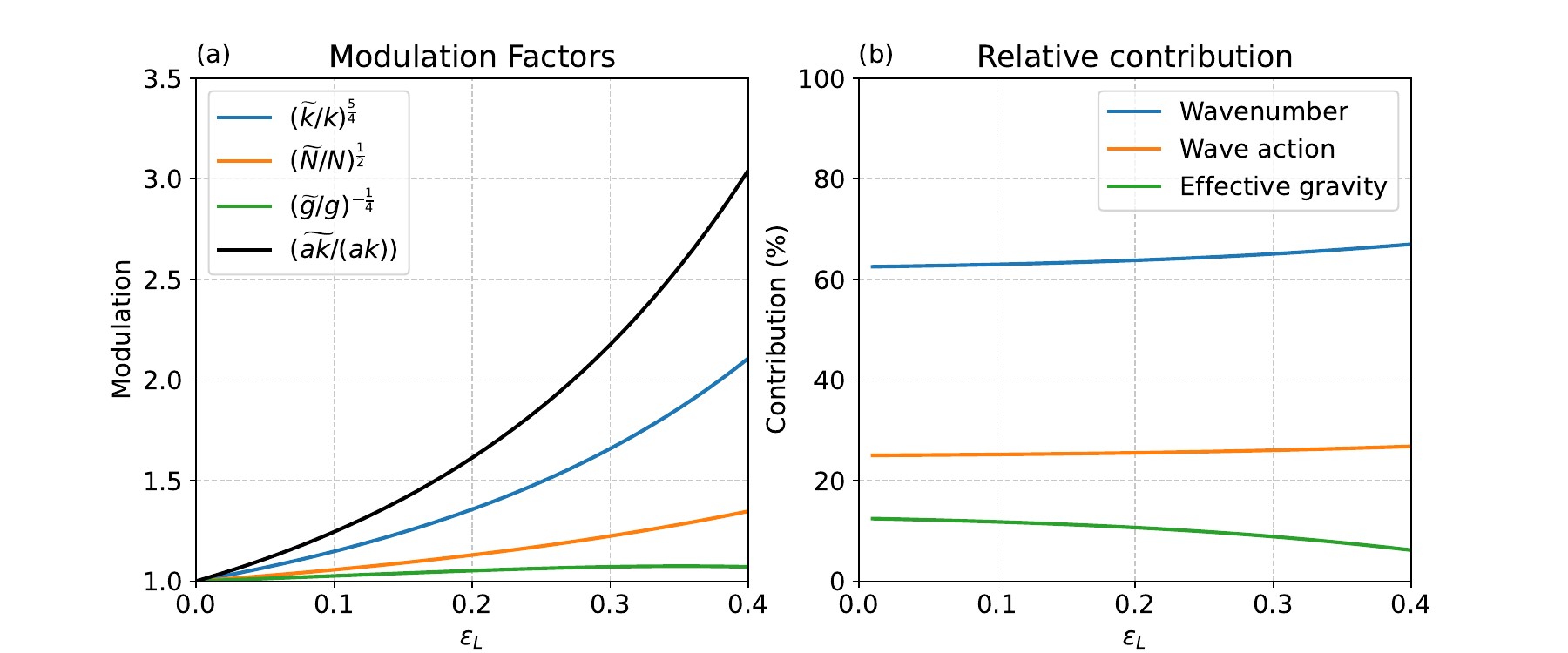}
\caption{
  Contributions of short-wave wavenumber, action, and effective gravity
  modulations to the steepness modulation, as function of the long-wave
  steepness $\varepsilon_L$.
  Panel (a) shows each modulation factor (colour) and their product (black),
  and panel (b) shows their relative contributions in percentage, calculated as
  the ratio of the logarithm of each modulation factor to the logarithm of the
  product of all modulations.
}
\label{fig:modulation_contributors}
\end{figure}

Finally, the short-wave steepness modulation follows from (\ref{eq:wave_amplitude_modulation}):

\begin{equation}
\label{eq:ak_modulation}
\dfrac{\widetilde{ak}}{ak} = 
  \left( \dfrac{\widetilde{k}}{k} \right)^{\frac{5}{4}}
  \left( \dfrac{\widetilde{N}}{N} \right)^{\frac{1}{2}}
  \left( \dfrac{\widetilde{g}}{g} \right)^{-\frac{1}{4}}
\end{equation}
The short-wave steepness modulation at the long-wave crests is thus largely
controlled by the wavenumber modulation (5/8, or, 62.5\%), followed by the
convergence of wave action (1/4, or, 25\%), and by least amount by the
effective gravity reduction (1/8, or, 12.5\%) (Fig. \ref{fig:modulation_contributors}).
(\ref{eq:ak_modulation}) and Fig. \ref{fig:modulation_contributors} demonstrate
that more than half of the steepness modulation can be captured by the
wavenumber modulation alone.
Further, considering the wavenumber and wave action modulations, while
neglecting the gravity modulation, captures $\approx$88\% of the steepness
modulation at the long-wave crests.
This result may be useful for the interpretation of remote sensing products
that resolve some but not all of the short-wave modulation effects.
Finally, the relative contributions of different modulation factors remain
mostly constant as $\varepsilon_L$ increases, with only a mild decrease of the
effective gravity modulation contribution, and mild increases of the
wavenumber and wave action modulation contributions.

\subsection{Limits to the wave action conservation}
\label{sec:wave_action_conservation}

Following the variational approach by \citet{whitham1965general},
\citet{bretherton1968wavetrains} showed that the wave action is conserved for
small-amplitude (linear) waves that propagate in slowly-varying media.
\citet{longuet1987propagation} correctly pointed out both
requirements--that both the medium and the short wavetrain must be
slowly-varying--for the wave action conservation to hold.
However, he incorrectly stated that there is no explicit restriction on the
steepness of the long wave, and that it appears necessary to only assume
small-amplitude short waves and significant scale separation ($k/k_L$).
The restriction on $\varepsilon_L$ becomes apparent when we recognize
that the long wave-induced orbital velocity scales with $\varepsilon_L$, and
that the divergence of this velocity is the dominant term in (\ref{eq:wavenumber})
and (\ref{eq:wave_action2}).
Here we show that in the general case both the homogeneity and the stationarity
of a short wave quantity $q$ depend on the long-wave steepness $\varepsilon_L$.

The homogeneity and stationarity of a scalar quantity $q$ can be expressed as
conditions on the instantaneous fractional rate of change of $q$ being much
smaller than the short-wave inverse scales:

\begin{equation}
  \frac{1}{q} \frac{\partial q}{\partial x} \ll k
  \label{eq:homogeneity_condition}
\end{equation}

\begin{equation}
  \frac{1}{q} \frac{\partial q}{\partial t} \ll \sigma
  \label{eq:stationarity_condition}
\end{equation}
The slowly-varying conditions on the medium (in our case, the long-wave orbital
velocity $U$) are then:
\begin{equation}
  \frac{1}{U} \frac{\partial U}{\partial x} \ll \widetilde{k},
\end{equation}

\begin{equation}
  \frac{1}{U} \frac{\partial U}{\partial t} \ll \widetilde{\sigma},
\end{equation}
respectively.
For a linear long wave, these conditions are equivalent to the scale separation
between the long and short waves:

\begin{equation}
  k_L \ll \widetilde{k}
  \label{eq:k_scale_separation}
\end{equation}

\begin{equation}
  \sigma_L \ll \widetilde{\sigma}
\end{equation}
\citet{bretherton1968wavetrains} also require that the medium is nearly uniform
over the vertical scales of motion of the short waves:

\begin{equation}
  \frac{1}{U} \frac{\partial U}{\partial z} \ll \widetilde{k}
\end{equation}
which, for a long deep water wave, is equivalent to (\ref{eq:k_scale_separation}).
Alternatively, if we use the second Froude number criterion following
\citet{ellingsen2014ship} and quantify the shear length
under the long wave surface as:

\begin{equation}
  l_S = \frac{g}{S^2} \approx \frac{g}{\varepsilon_L^2 \omega_L^2}
\end{equation}
we arrive to an additional requirement:

\begin{equation}
  \frac{k}{k_L} \gg \varepsilon_L^2
\end{equation}
which relates the scale separation to the long-wave steepness.

\begin{figure}
\centering
\includegraphics[width=\textwidth]{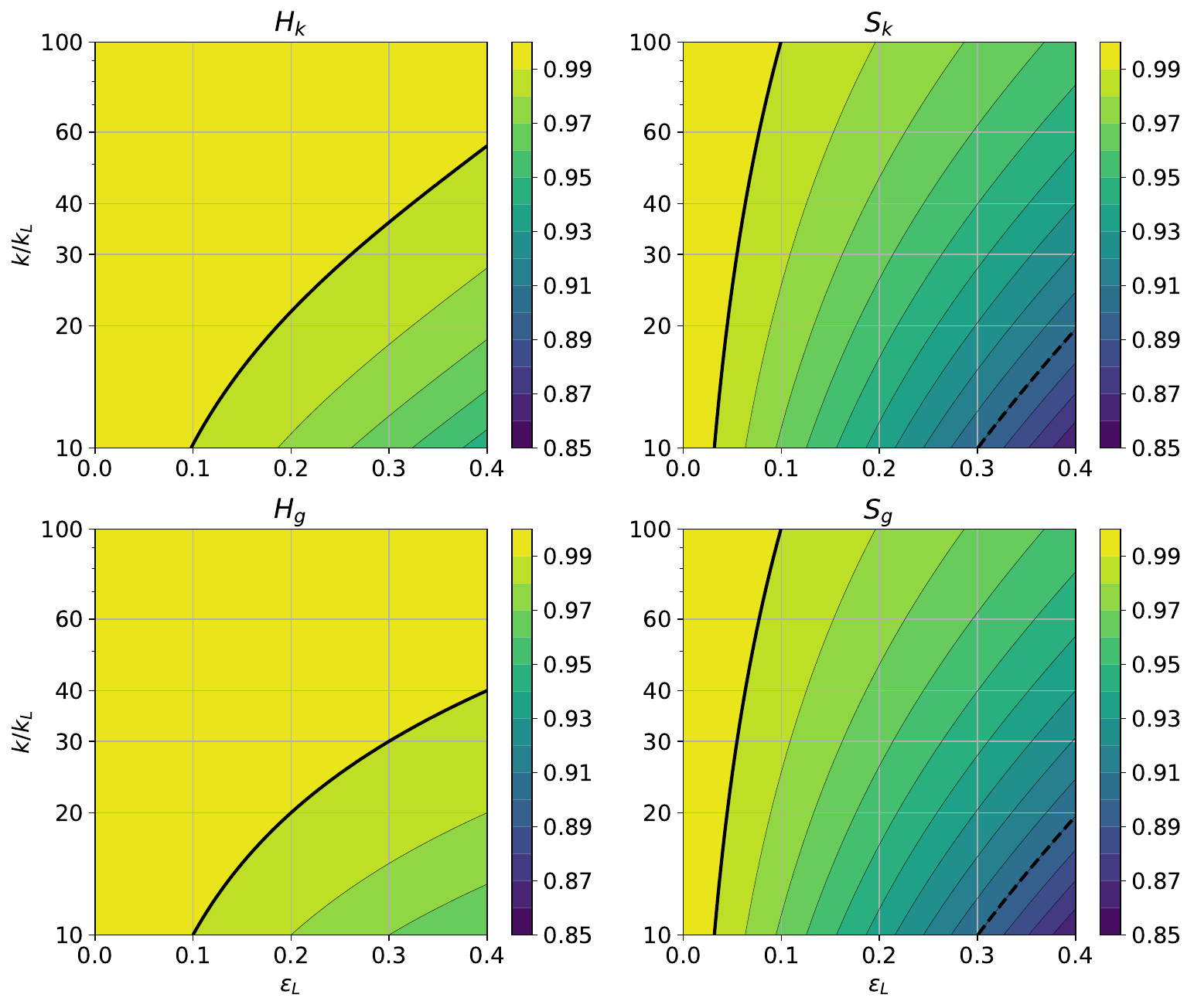}
\caption{
Homogeneity (left) and stationarity (right) of the short-wave wavenumber (top)
and effective gravity (bottom) as a function of the long-wave steepness
$\varepsilon_L$ and the wavenumber ratio $k/k_L$, based on the linearised
solutions (\ref{eq:homogeneity_k}-\ref{eq:stationarity_g}).
Solid and dashed lines highlight the 0.99 and 0.90 values, respectively.
}
\label{fig:homogeneity_stationarity_analytical}
\end{figure}

Now, to address the variation of short-wave properties,
we apply (\ref{eq:homogeneity_condition})-(\ref{eq:stationarity_condition}) to
the long-wave phase dependent wavenumber, gravitational acceleration, and
wave action, and define the homogeneity and stationarity as:

\begin{equation}
  H_q = 1 - \left| \frac{1}{qk} \frac{\partial q}{\partial x} \right|
\end{equation}

\begin{equation}
  S_q = 1 - \left| \frac{1}{q\sigma} \frac{\partial q}{\partial t} \right|
\end{equation}
With the above definitions, we consider $q$ to be homogeneous and stationary as
$H_q \rightarrow 1$ and $S_q \rightarrow 1$, respectively.

For the linearised modulation solutions (\ref{eq:k_short_exact}),
(\ref{eq:gravity_modulation_linear_no_curvature_mean_level}), and
(\ref{eq:wave_action_modulation}), the respective homogeneity expressions are:

\begin{equation}
  H_{k,N} = 1 - \frac{k_L}{k} \left| \frac{\varepsilon_L \sin{\psi}}{\left(1 + \varepsilon_L \cos{\psi}\right)^2} \right|
  \label{eq:homogeneity_k}
\end{equation}

\begin{equation}
  H_{g} = 1 - \frac{k_L}{k} \left| \frac{\varepsilon_L \sin{\psi}}{\left(1 - \varepsilon_L^2 \cos^2{\psi}\right)} \right|
  \label{eq:homogeneity_g}
\end{equation}

\begin{equation}
  S_{k,N} = 1 - \frac{\sigma_L}{\sigma} \left| \frac{\varepsilon_L \sin{\psi}}{\left(1 + \varepsilon_L \cos{\psi}\right) \sqrt{1 - \varepsilon_L^2 \cos^2{\psi}}} \right|
  \label{eq:stationarity_k}
\end{equation}

\begin{equation}
  S_{g} = 1 - \frac{\sigma_L}{\sigma} \left| \frac{\varepsilon_L \sin{\psi}}{\left(1 - \varepsilon_L \cos{\psi}\right) \sqrt{1 - \varepsilon_L^2 \cos^2{\psi}}} \right|
  \label{eq:stationarity_g}
\end{equation}
The steady, linearised solutions thus depend on both the wavenumber ratio
$k_L/k$ (homogeneity) or the frequency ratio $\sigma_L/\sigma$ (stationarity)
and the long-wave steepness $\varepsilon_L$.

Fig. \ref{fig:homogeneity_stationarity_analytical} shows the homogeneities and
stationarities of the short-wave wavenumber and effective gravity based on
the steady solutions and their derived criteria as minima of
(\ref{eq:homogeneity_condition})-(\ref{eq:stationarity_condition}).
Stationarity is a stronger requirement than homogeneity.
At the lowest scale separation considered ($k/k_L = 10$), short-wave wavenumber,
action, and effective gravity are all strongly homogeneous ($H_{k,N,g} > 0.99$)
for $\varepsilon_L < 0.1$.
However, the solutions are only strongly stationary ($S_{k,N,g} > 0.99$) at a
high scale separation of $k/k_L = 100$ for $\varepsilon_L < 0.1$.
If we consider weakly stationary conditions as $S_{k,N,g} > 0.9$, then the
solutions satisfy the requirements for the wave action conservation for
$\varepsilon_L < 0.3$ at $k/k_L = 10$, and for $\varepsilon_L < 0.4$ at
$k/k_L > 20$.
Although \citet{bretherton1968wavetrains} state that the wave action balance
is valid to $\mathcal{O}(k_L/k)$, applying the general criteria
(\ref{eq:homogeneity_condition})-(\ref{eq:stationarity_condition}) yields an
error of $\mathcal{O}(\varepsilon_L k_L/k)$.
We will revisit these criteria based on the numerical solutions of nonlinear
wave action and crest conservation equations in the next section.
For the asymptotic limits of wave action conservation in terms of the wave
short-wave steepness, see Appendix \ref{appendix:wave_action_conservation}.

\section{Numerical solutions}
\label{section:numerical_solutions}

\subsection{Model description}

To quantify the contribution of the nonlinear terms in (\ref{eq:wavenumber})
and (\ref{eq:wave_action2}), and to evaluate the steadiness assumption of the
analytical solutions, we now proceed to numerically integrate the full set of
crest and action conservation equations.
(\ref{eq:wavenumber}) and (\ref{eq:wave_action2}) are discretised using
second-order central finite difference in space and integrated in time using the
fourth-order Runge-Kutta method \citep{butcher1996runge}.
The space is divided into 128 grid points.
This is effectively the same equation set and numerical configuration as that of
\citet{peureux2021unsteady}, except that their long-wave orbital velocities are
evaluated at $z=0$ instead of $z=\eta_L$, thus neglecting the Stokes drift
induced by long waves on short-wave groups
\citep{stokes1847,van2018stokes,monismith2020stokes}.
Their gravity modulation also does not consider the nonlinear, Lagrangian, or
slope effects, however, this difference does not qualitatively affect the results.
Another difference is the choice of the numerical scheme for spatial differences,
which in their case is the more sophisticated MUSCL4 scheme \citep{kurganov2000new}.
Although the second-order central finite difference is not appropriate for many
numerical problems, in our case it is sufficient because it is conservative and
the fields that we compute derivatives of are smooth and continuous.
The maximum relative error of the centred finite difference in this model
is $\approx 0.066\%$.
The wavenumber and wave action are conservative over time within
$\mathcal{O}(10^{-7})$ and $\mathcal{O}(10^{-5})$ relative error, respectively.
We show in the next subsection that our numerical solutions for infinite long-wave
trains are qualitatively equivalent to those of \citet{peureux2021unsteady}.

The numerical equations here are integrated in a fixed reference frame with
periodic boundary conditions, rather than that moving with the long-wave phase speed.
Either approach produces equivalent modulation results, however a fixed
reference frame allows for a more intuitive interpretation of the results.
This difference is important to keep in mind when comparing the results
of \citet{peureux2021unsteady} and those presented here; in their Fig. 1 the
long-wave phase is fixed and the short waves are moving leftward
with the speed of $C_{pL} - C_g - U$ (neglecting group speed inhomogeneity),
whereas in the figures that follow, the long wave is moving rightward with its
phase speed $C_{pL}$ and the short waves (that is, their action) are moving
rightward with the speed of $C_g + U$ (neglecting group speed inhomogeneity).
The crest and action conservation equations are integrated in the curvilinear
coordinate system in which the horizontal coordinate follows the long-wave
surface, akin to \citet{zhang1990evolution}.

\subsection{Comparison with analytical solutions}
\label{subsection:comparison_with_analytical_solutions}

\begin{figure}
\centering
\includegraphics[width=\textwidth]{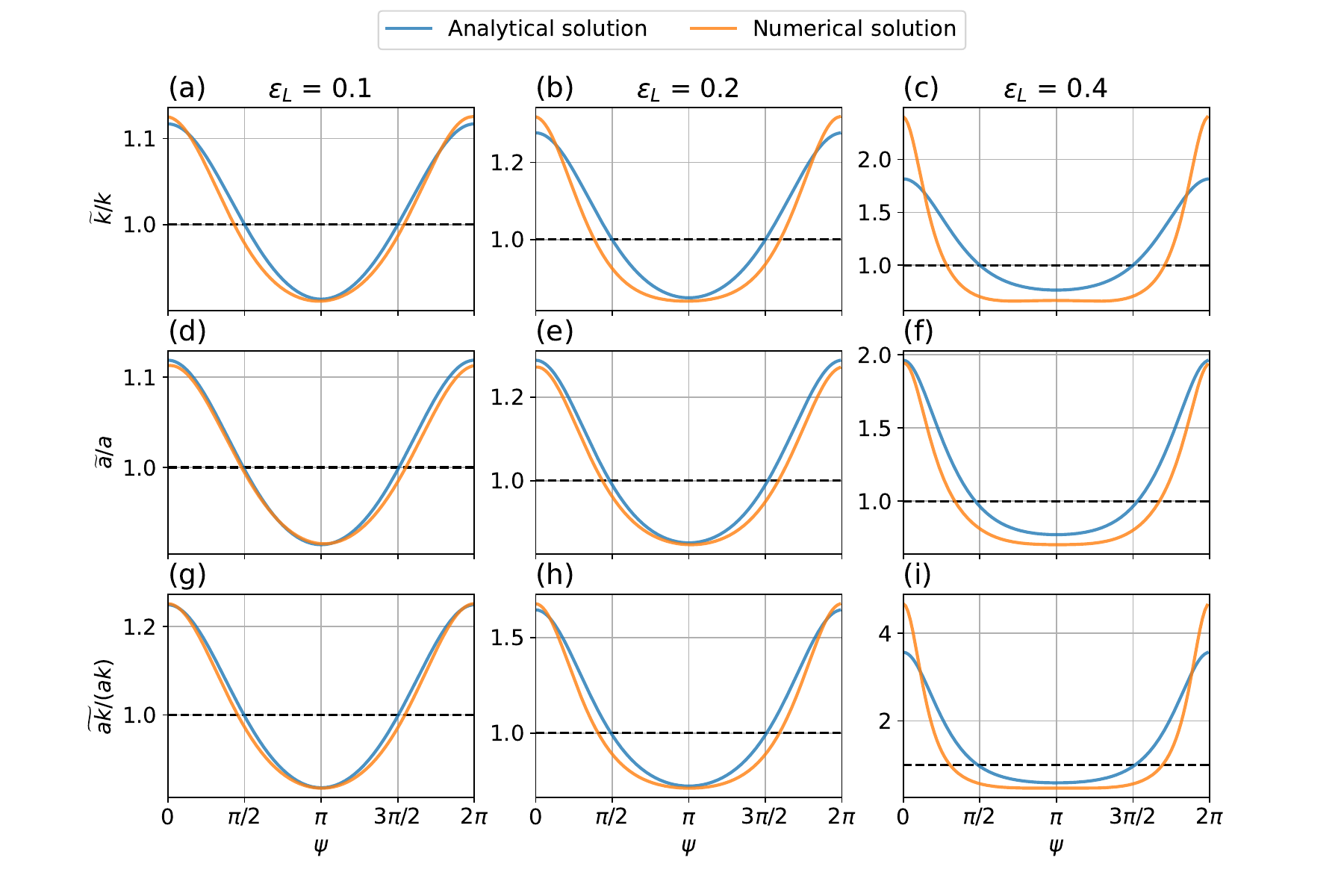}
\caption{
  Comparison of numerical solutions (orange) of wavenumber (top), amplitude
  (middle), and steepness (bottom) modulation with their analytical solutions
  (blue), for $\varepsilon_L$ = 0.1, 0.2, and 0.4.
}
\label{fig:numerical_solutions}
\end{figure}

The first set of simulations is performed for the long-wave steepness of
$\varepsilon_L = 0.1$, $0.2$, and $0.4$ (Fig. \ref{fig:numerical_solutions}).
The long-waves are initialised from rest ($a_L = 0$) and gradually ramped up
to their target steepness over 5 long-wave periods to allow for a gentle
increase of the long-wave forcing on the short waves
(the importance of which we discuss in more detail in the next subsection).
At $\varepsilon_L = 0.1$, the numerical solutions for the wavenumber, amplitude,
and steepness modulation are all remarkably similar to the analytical solutions
(Fig. \ref{fig:numerical_solutions}a,d,g).
This suggests that for low $\varepsilon_L$ the steady analytical solution
is a reasonable approximation for the nonlinear crest-action solutions.
The numerical solutions diverge more notably from the analytical ones at
moderate $\varepsilon_L = 0.2$ (Fig. \ref{fig:numerical_solutions}b,e,h),
and are considerably different at very high $\varepsilon_L = 0.4$
(Fig. \ref{fig:numerical_solutions}c,f,i).
These differences suggest that at high $\varepsilon_L$ the inhomogeneity of
gravitational acceleration and group speed, otherwise ignored in the steady
solutions, become important.

\begin{figure}
\centering
\includegraphics[width=0.8\textwidth]{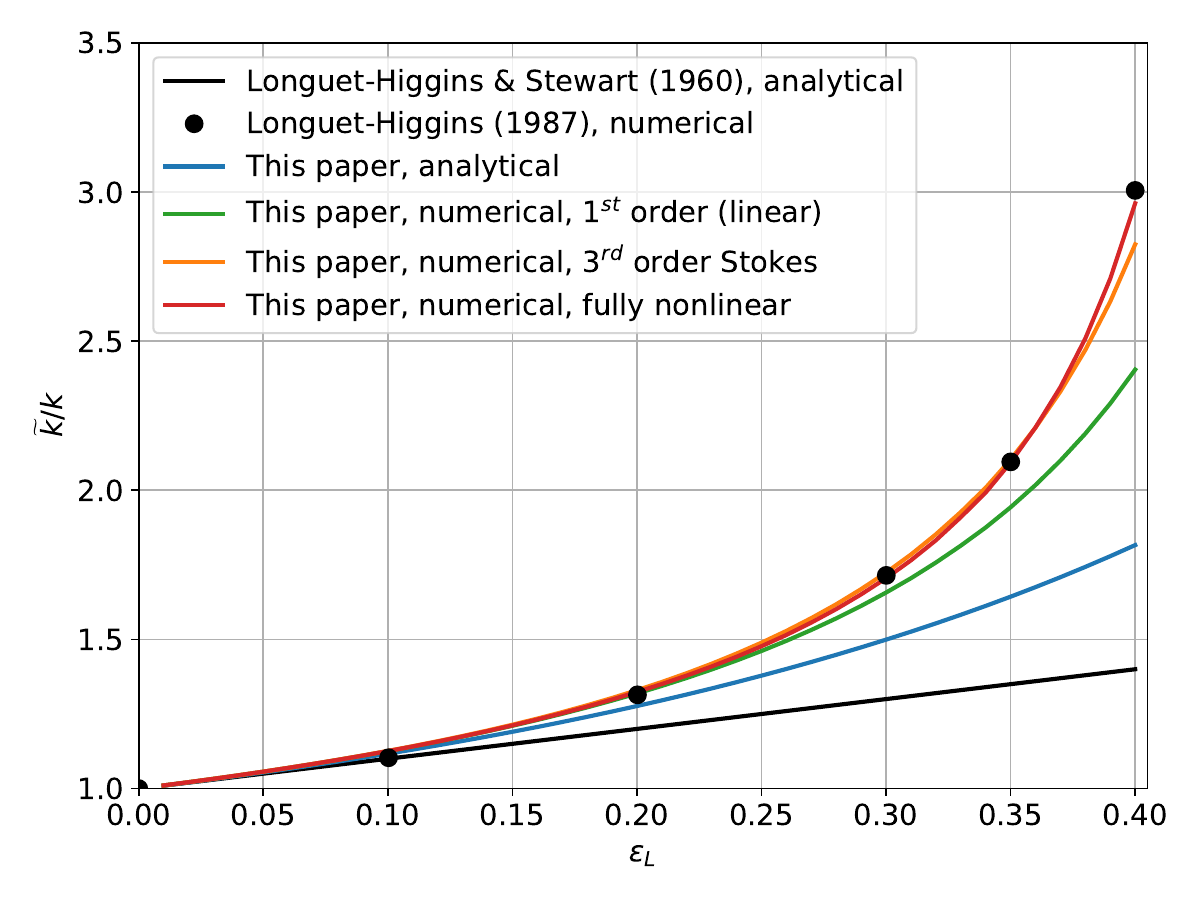}
\caption{
  Maximum wavenumber modulation as function of long-wave steepness $\varepsilon_L$.
  Black line and circles are for the analytical and numerical solutions from
  \citet{longuet1960changes} and \citet{longuet1987propagation}, respectively.
  Blue line is based on the analytical solutions from this paper.
  Green and orange lines are for the numerical solutions from this paper
  using linear and third-order Stokes approximations of long waves, respectively.
  Red line is for the fully-nonlinear gravity wave based on \citet{clamond2018accurate}.
}
\label{fig:numerical_vs_analytical_k}
\end{figure}

\begin{figure}
\centering
\includegraphics[width=0.8\textwidth]{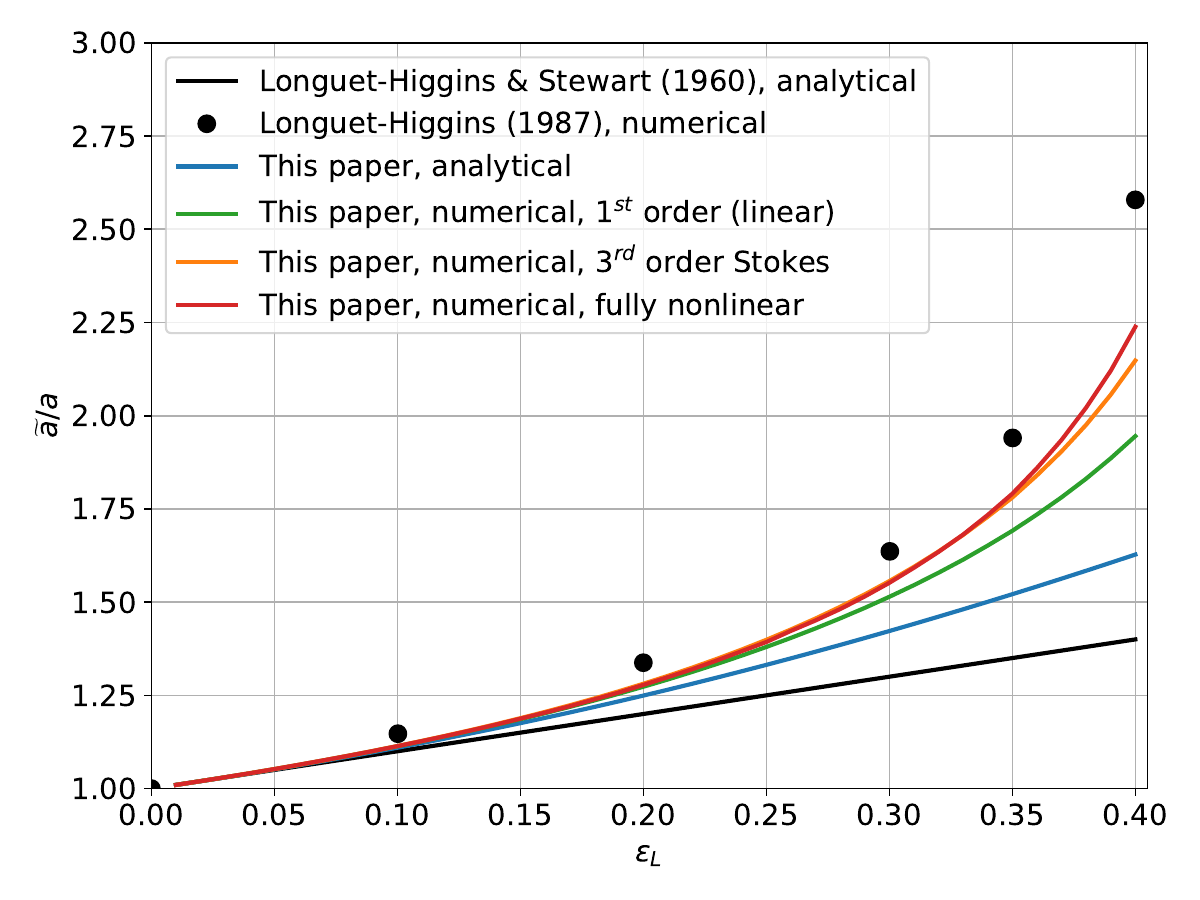}
\caption{
  As Fig. \ref{fig:numerical_vs_analytical_k} but for the amplitude modulation.
}
\label{fig:numerical_vs_analytical_a}
\end{figure}

\begin{figure}
\centering
\includegraphics[width=0.8\textwidth]{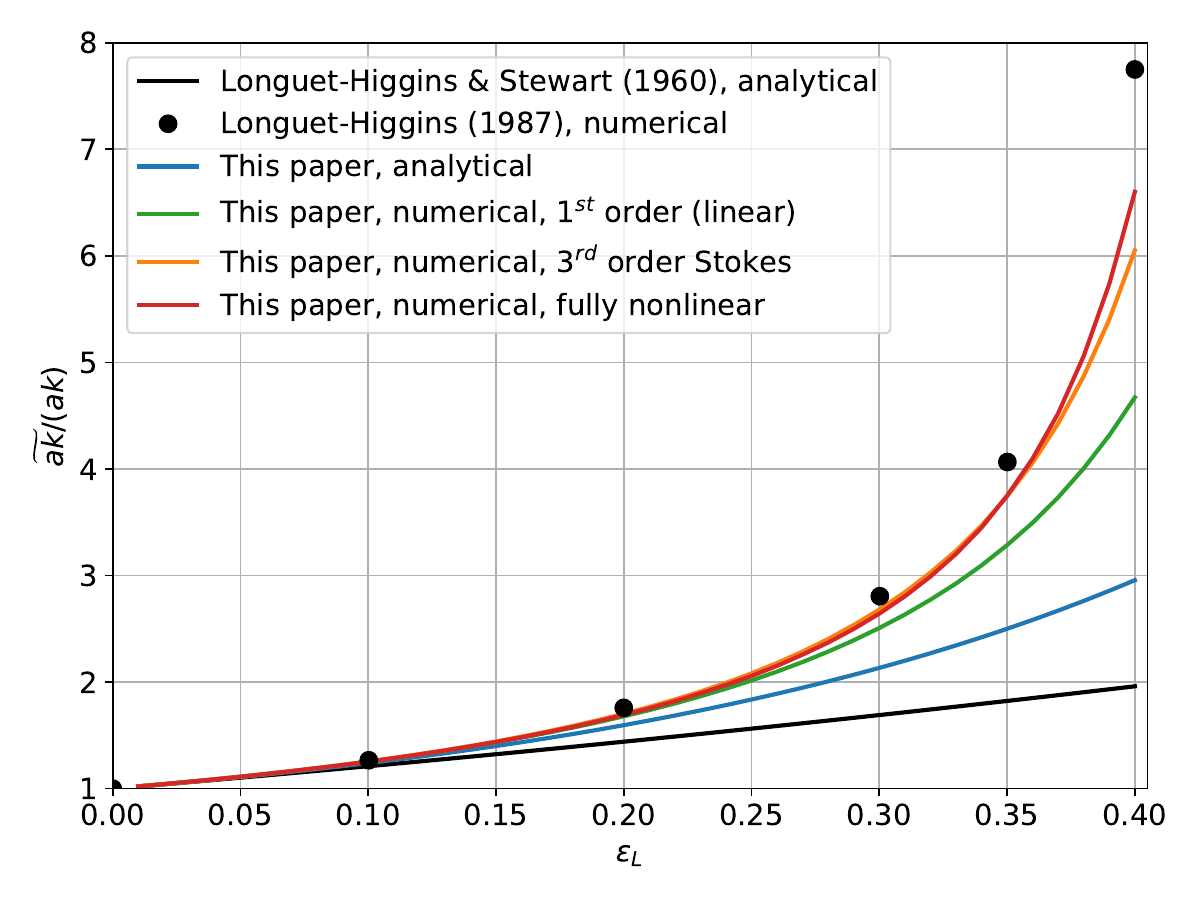}
\caption{
  As Fig. \ref{fig:numerical_vs_analytical_k} but for the steepness modulation.
}
\label{fig:numerical_vs_analytical_ak}
\end{figure}

The maximum modulations of the short-wave wavenumber, amplitude, and steepness
from the analytical and numerical solutions, are shown on Figs.
\ref{fig:numerical_vs_analytical_k},
\ref{fig:numerical_vs_analytical_a}, and
\ref{fig:numerical_vs_analytical_ak}, respectively, as function of the long-wave
steepness $\varepsilon_L$.
Overall, the numerical solutions, based on the linear and the third-order Stokes
long waves alike, begin to diverge from the analytical solution at
$\varepsilon_L \approx 0.15$.
The numerical solution using the third-order Stokes long wave is quantitatively
very similar to the fully-nonlinear solutions of \citet{longuet1987propagation}
in wavenumber modulation, however with somewhat lower amplitude and steepness
modulations.
Using the velocity and gravity properties of a fully-nonlinear wave, based on
the approach by \citet{clamond2018accurate}, the numerical solutions diverge
from those of the third-order Stokes wave only at very large steepnesses
($\varepsilon_L > 0.3$).
Using a fully-nonlinear form of the long-wave thus does not collapse the
difference between the solutions here and those of \citet{longuet1987propagation}.
For example, at $\varepsilon_L = 0.4$, the steepness modulation in our solution
is 6.6, whereas \citet{longuet1987propagation} finds it to be 7.8.
A possible reason for this difference is that the solutions by
\citet{longuet1987propagation} are stationary (see his \S2), whereas in our
numerical model we solve the prognostic nonlinear equations for the short-wave
wavenumber and action.
As we will see in the next section, and as was suggested by \citet{peureux2021unsteady},
they are often, but not always, mostly steady, and can be strongly unsteady in
special cases.
Nevertheless, the overall steepness modulation between the two numerical
approaches are remarkably similar, and either solution would cause the
short-waves of moderate steepness to break \citep{banner1993wave}.
Specifically, where the two modulation solutions differ ($\varepsilon_L \gtrsim 0.3$),
the steepness modulation approaches and exceeds 3.
Tripling the short-wave steepness of even $\varepsilon = 0.1$ would bring the
wave train near the observed breaking thresholds of $\varepsilon \gtrsim 0.32$
\citep{perlin2013breaking}, although, recent measurements show evidence of
significantly higher breaking thresholds
\citep{toffoli2010maximum,mcallister2024three}, and even exceeding the
geometric Stokes limit of 0.44.
Nevertheless, in the cases where the short-waves would survive the
modulation and persist over multiple long-wave periods, the two solutions are
effectively equivalent.

\begin{figure}
\centering
\includegraphics[width=\textwidth]{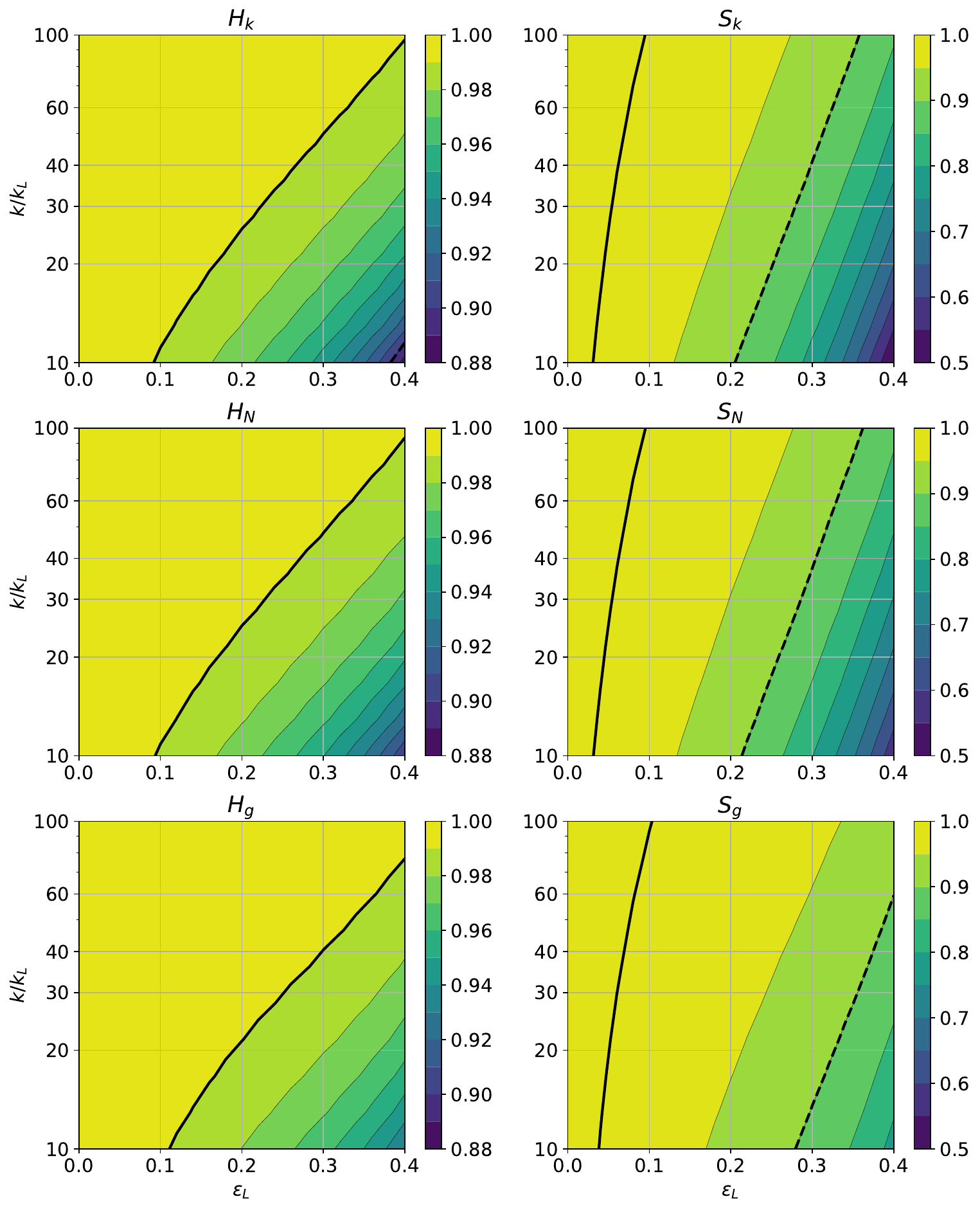}
\caption{
Homogeneity (left) and stationarity (right) of the short-wave wavenumber (top),
action (middle), and effective gravity (bottom) as a function of the long-wave
steepness $\varepsilon_L$ and the wavenumber ratio $k/k_L$, based on the
numerical solutions of the full wave crest and action conservation equations.
Note that the color ranges are different than those in Fig. \ref{fig:homogeneity_stationarity_analytical}.
}
\label{fig:homogeneity_stationarity_numerical}
\end{figure}

As the numerical solutions are for the nonlinear wave action and crest conservation
equations, it is important to re-evaluate the homogeneity and stationarity
requirements for the wave action conservation to remain valid.
Fig. \ref{fig:homogeneity_stationarity_numerical} shows the homogeneities
and stationarities of the short-wave wavenumber, action, and effective gravity
as function of the long-wave steepness $\varepsilon_L$ and the wavenumber ratio
$k/k_L$, based on the numerical simulations.
Although homogeneity is slightly lower ($H_{k,N,g} \approx 0.92$ for $k/k_L = 10$
and $\varepsilon_L = 0.4$) compared to the analytical solutions,
the stationarity is significantly reduced.
Following the same criteria for strong ($S_{k,N,g} > 0.99$) and weak
($S_{k,N,g} > 0.9$) stationarity as in \S\ref{sec:wave_action_conservation},
the numerical solutions are only weakly stationary for $\varepsilon_L < 0.21$
at $k/k_L = 10$, and for $\varepsilon_L < 0.36$ at $k/k_L = 100$.
This provides a quantitative guidance on when the short-wave action can be
considered to be conserved in the presence of longer waves.

\subsection{Unsteady growth in Pereux et al. (2021)}
\label{subsection:unsteady_growth}

\begin{figure}
\centering
\includegraphics[width=\textwidth]{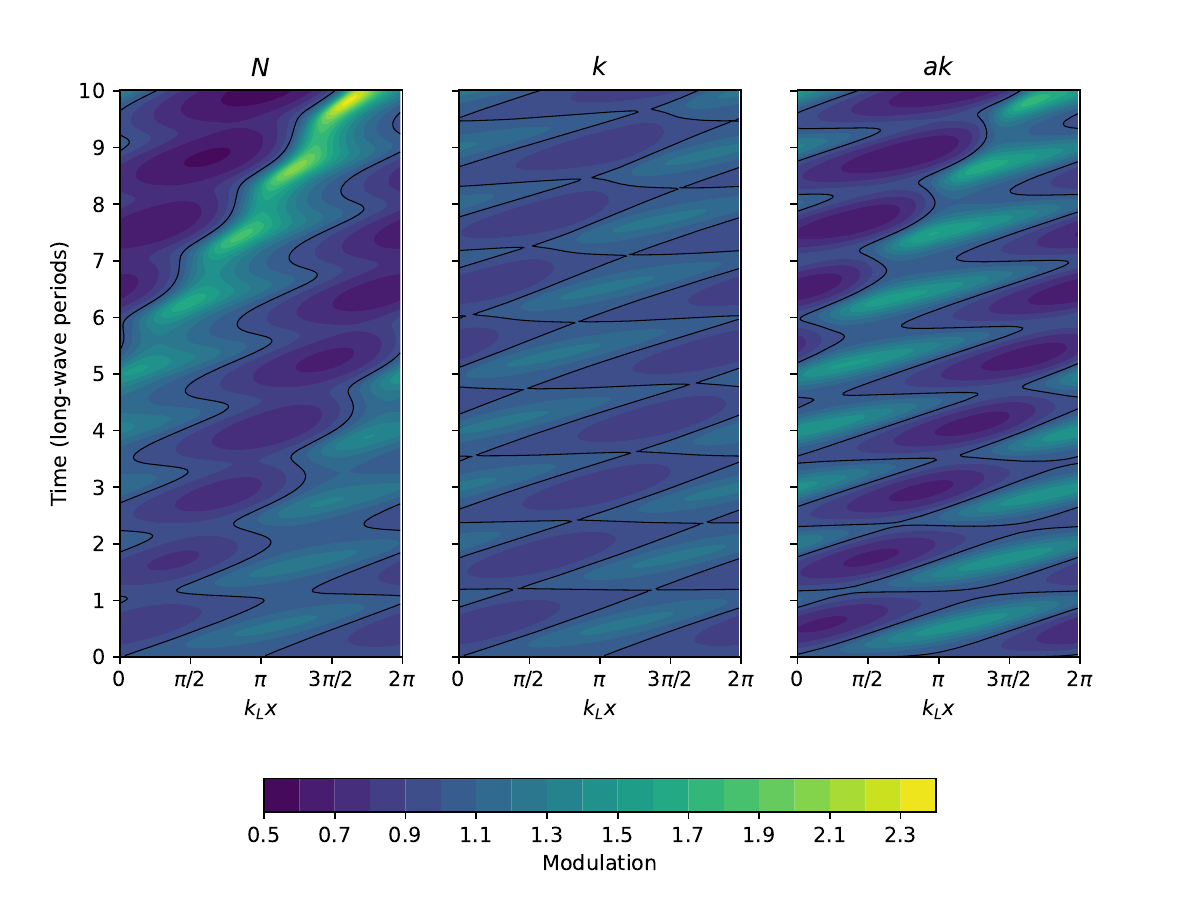}
\caption{
  Change in the short-wave action (left), wavenumber (middle), and steepness (right)
  relative to their initial values as function of initial long-wave phase and time.
  $k_L = 1$, $k = 10$, $\varepsilon_L = 0.1$.
  The black contour follows the value of 1, which corresponds to no change
  from the initial values.
}
\label{fig:modulation_3panel_infinite}
\end{figure}

The key result of \citet{peureux2021unsteady} is that of unsteady steepening of
short waves when a homogeneous field of short waves is forced by an infinite
long-wave train.
The steepening is driven by the progressive increase in the action of short
waves (as opposed to their wavenumber), and is centred on the short-wave group
that begins its journey in the trough of the long wave.
This is because the unmodulated short waves that begin their journey in the
trough first enter the convergence region before reaching the divergence region.
Inversely, the short waves that begin around the crest of the long-wave first
enter the divergence region and will thus have their action and wavenumber
decreased.
We reproduce this behaviour in Fig. \ref{fig:modulation_3panel_infinite}, which
shows the evolution of the short-wave action, wavenumber, and steepness
modulation (here defined as the change of a quantity relative to its initial value).
The long waves have $\varepsilon_L = 0.1$ and $k_L = 1$, and the short waves
have $k = 10$.
After 10 long-wave periods, the short-wave action is approximately doubled,
consistent with \citet{peureux2021unsteady}.
Another important property of this solution is that the short-wave group that
experiences peak amplification (attenuation) are not locked-in to the long-wave
crests (troughs), as they are in the prior steady solutions
\citep{longuet1960changes,longuet1987propagation,zhang1990evolution}.

Why does this unsteady growth of short-wave action occur?
\citet{peureux2021unsteady} explain that the change in advection velocity due to
the modulation of the short-wave group speed yields an additional amplification
of the short-wave action.
That is, the inhomogeneity term $N \partial C_g / \partial x$ is
responsible for the instability.
However, the instability only occurs if the short waves are initialised as a
uniform field.
If they are initialised instead as a periodic function of the long-wave phase
such that their wavenumber (and optionally, action) is higher on the crests
and lower in the troughs, akin to the prior steady solutions, the instability
vanishes and the short-wave modulation pattern locks-in to the long-wave
crests.

\begin{figure}
\centering
\includegraphics[width=\textwidth]{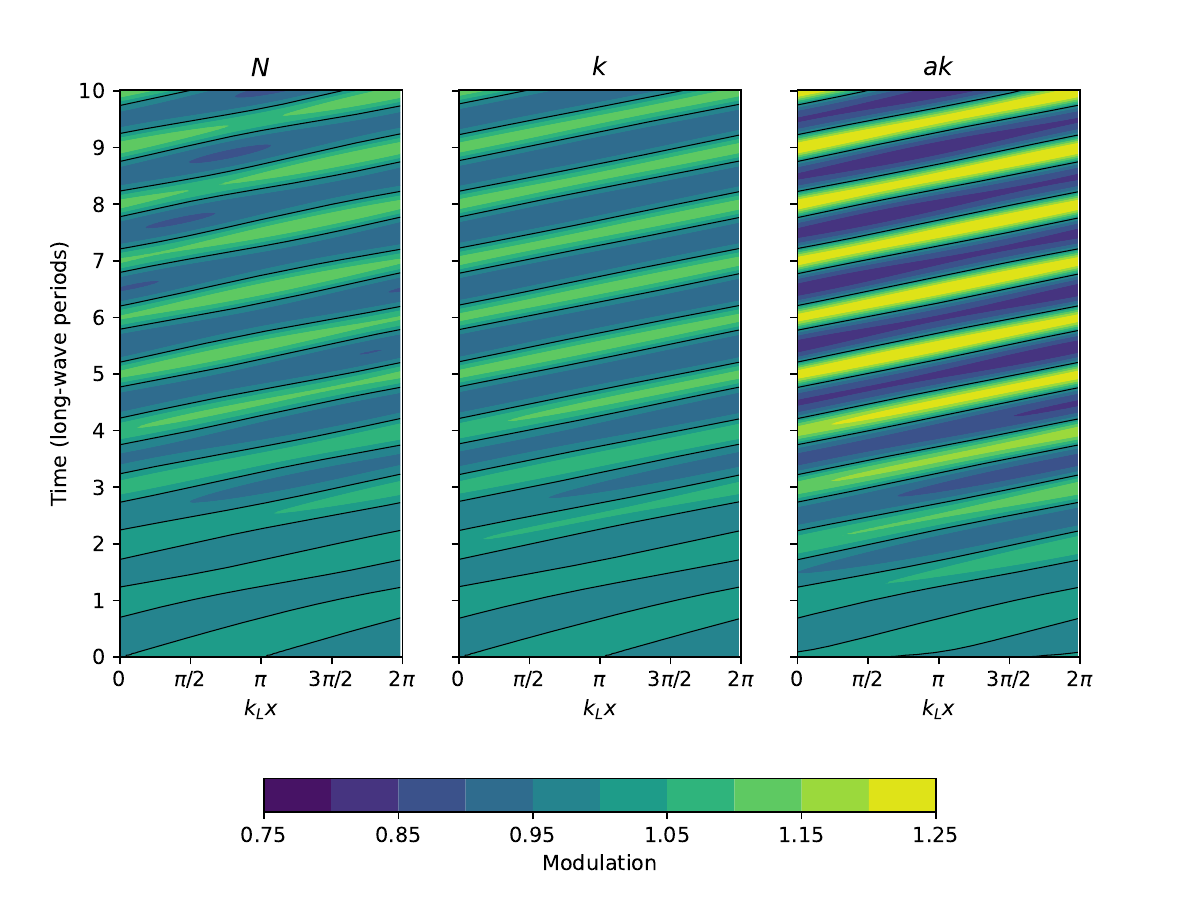}
\caption{
  Same as Fig. \ref{fig:modulation_3panel_infinite} but with a linear ramp
  applied to $a_L$ during the initial five long-wave periods.
  Notice the change in the colour range.
}
\label{fig:modulation_3panel_ramp}
\end{figure}

Why is it, then, that the solution for short-wave modulation is not only
quantitatively, but also qualitatively, different depending on the short-wave
initial conditions?
It may at first seem that there is something special about the uniform initial
conditions that makes its solutions grow unsteadily and indefinitely.
\citet{peureux2021unsteady} described this scenario as "...the sudden appearance
of a long wave perturbation in the middle of a homogeneous short wave field".
However, it is not obvious whether such a scenario is common as waves tend to
disperse into groups, and the time scales of incoming swell fronts are much
longer than those of short-wave generation by wind.
Exceptional situations with such a sudden onset of swell may occur
around complex coastlines or in shadows behind islands, where wind-generated
ripples may propagate from an otherwise calm sea state into the path of a
crossing swell.
Within the limits of the modulation theory used here, such short waves could
quickly steepen and break.

\begin{figure}
  \centering
  \includegraphics[width=0.8\textwidth]{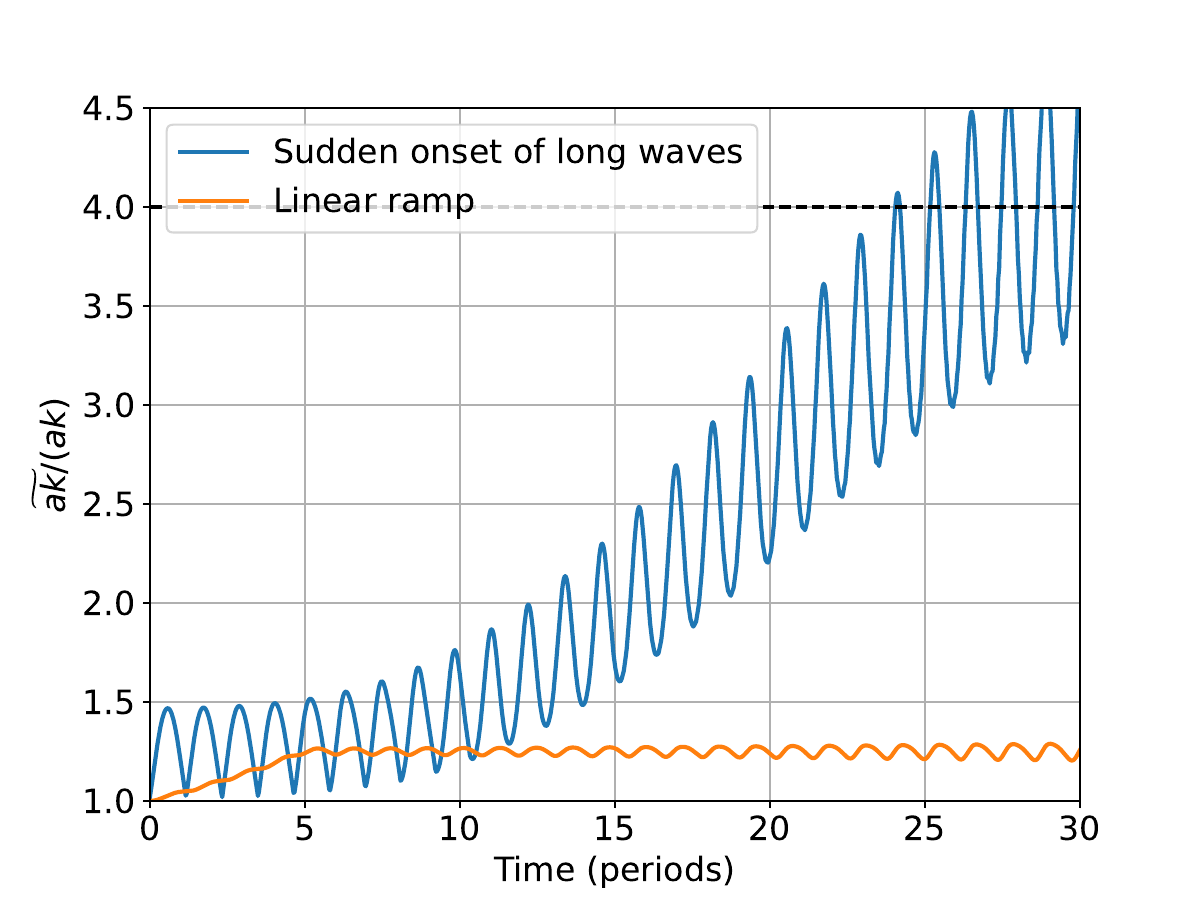}
  \caption{
    Maximum short-wave steepness modulation as function of time in case of
    infinite long-wave train case (blue) and the same case but with a linear
    ramp during the initial five long-wave periods (orange). The dashed black
    line corresponds to the short-wave steepness of 0.4 at which most waves are
    expected to break.
  }
  \label{fig:unsteady_growth_timeseries}
\end{figure}

\begin{figure}
  \centering
  \includegraphics[width=\textwidth]{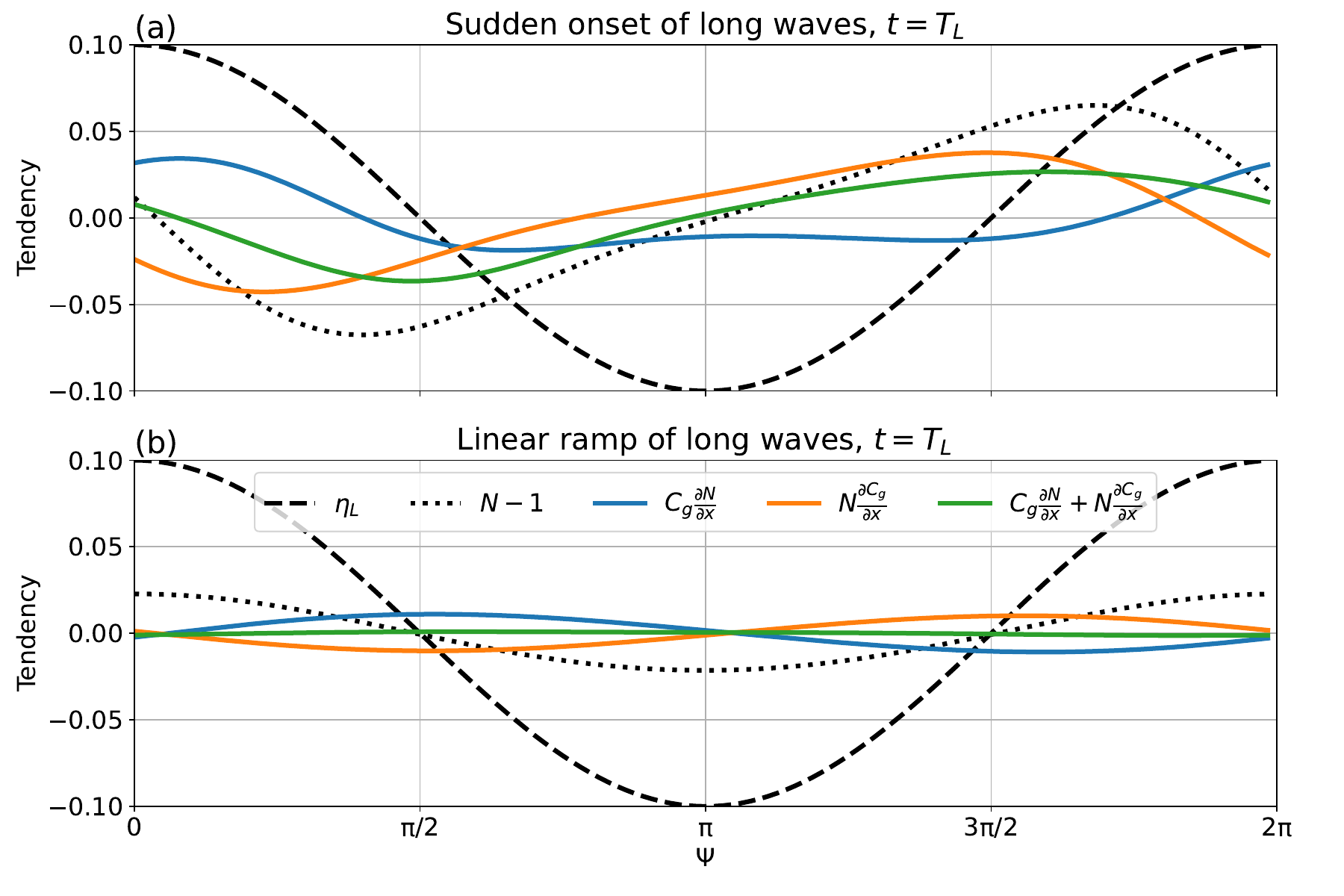}
  \caption{
    Propagation ($C_g \frac{\partial N}{\partial x}$) (blue) and inhomogeneity
    ($N \frac{\partial C_g}{\partial x}$) (orange) tendencies and their sum
    (green) for the short-wave action after one long-wave period, in the case of
    (a) infinite long-wave train and (b) linear ramp of long-wave amplitude.
    Dashed line is the long-wave elevation and the dotted line is the short-wave
    action modulation minus 1.
  }
  \label{fig:inhomogeneity_tendencies}
\end{figure}

To test whether the sudden onset of long-wave forcing causes the unsteady growth,
we repeat the numerical simulation of Fig.
\ref{fig:modulation_3panel_infinite}, except for a linear ramp applied to $a_L$
(and consequently, the other long-wave properties including the orbital
velocities and the local steepness) for the first five periods:
$a_L(t) = min(a_L, a_L t / \left(5 T_L\right))$, where $T_L$ is the long-wave period.
This result is shown in Fig. \ref{fig:modulation_3panel_ramp}.
Note that the linear ramp is strictly a numerical modeling technique to test
the sensitivity of the solution to the sudden forcing at the beginning of the
simulation, and is not intended to represent a physical process.
The key distinction that arises with the introduction of the ramp is that the
modulation pattern of short waves is now locked-in to the long-wave crests,
and their indefinite steepening vanishes.
The stabilisation effect of the long-wave ramp is more readily assessed in
Fig. \ref{fig:unsteady_growth_timeseries}, which shows the maximum short-wave
steepness modulation as function of time over 30 long-wave periods, for the
infinite long-wave train case and the case of applying a linear ramp.
Comparing the wave action propagation ($C_g \partial N / \partial x$) and
inhomogeneity ($N \partial C_g / \partial x$) tendencies reveals that
after only one long-wave period, the propagation and inhomogeneity tendencies
are unbalanced and amplify the steepening of short waves
(Fig. \ref{fig:inhomogeneity_tendencies}).
\citet{peureux2021unsteady} correctly pointed out that the unsteady growth of
short-wave steepness occurs due to the resonance of the inhomogeneity
tendency with the short-wave action perturbation.
However, in the case of a gentle ramp-up of the long-wave forcing, the
propagation and inhomogeneity tendencies balance each other and allow
the short-wave steepening to lock-in to the long-wave crests.

\subsection{Modulation by long waves of varying amplitude}
\label{subsection:wave_groups}

\begin{figure}
\centering
\includegraphics[width=\textwidth]{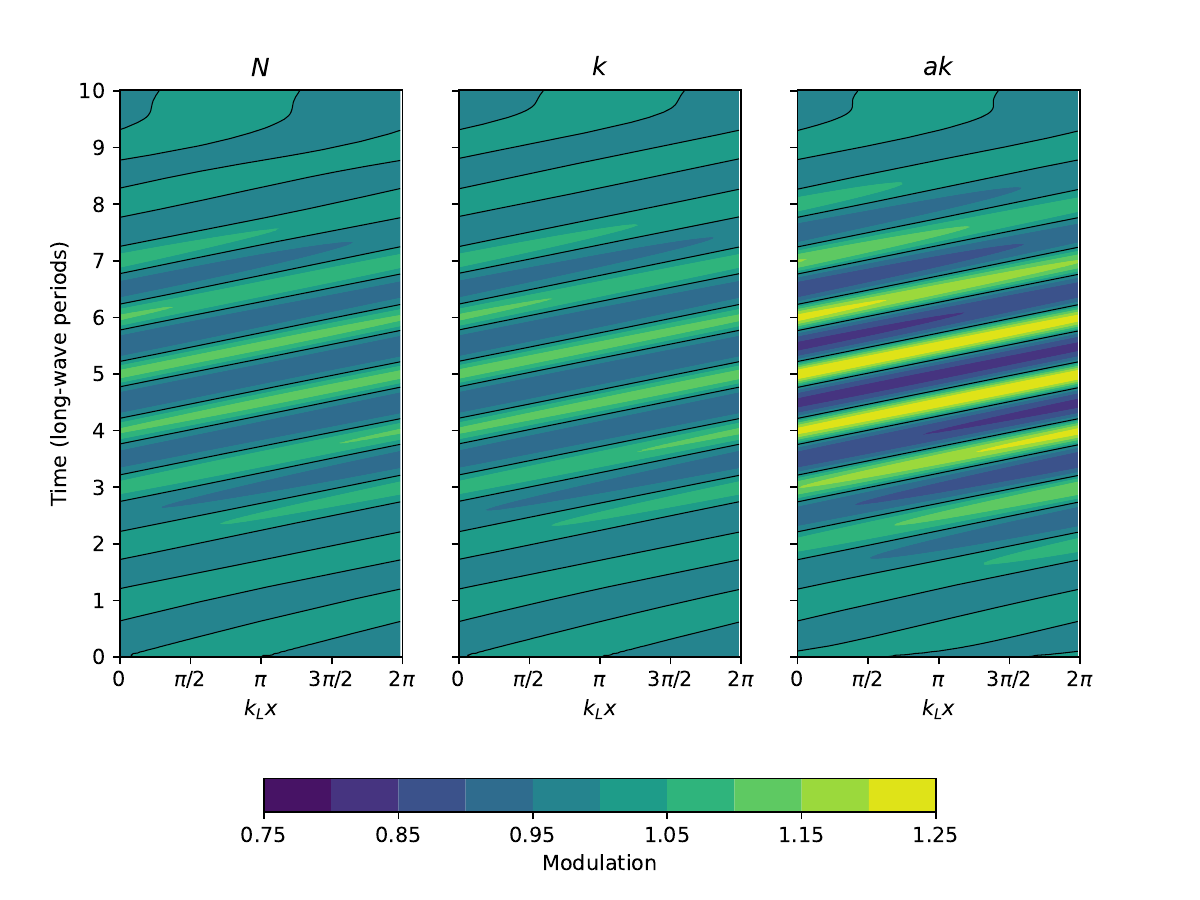}
\caption{
  Same as Fig. \ref{fig:modulation_3panel_infinite} but for a long-wave group,
  causing the long-wave amplitude to gradually increase and peak at $a_L = 0.1$
  after five long-wave periods, and then conversely decay back to a calm sea state.
}
\label{fig:modulation_3panel_groups}
\end{figure}

We now proceed to examine the numerical solutions of short-wave modulation by
long waves whose amplitude varies with time.
The amplitude variation aims to mimic the behaviour of long-wave groups, a
phenomenon that is ubiquitous in the ocean due to the dispersive nature of
surface gravity waves, as well as due to the fact that wind-generated waves
evolve to have broad-banded spectra.
For simplicity of implementation and interpretation, 
the wave group here is simulated as a non-dispersive train of long waves whose
amplitude scales with $\sin[\pi t / (n T_L)]$, where $n$ is the number of
long waves in the group.
The elevation of the long waves thus gradually ramps up from rest to $a_L$ at
$t = n T_L / 2$, after which it gradually decreases toward zero.
In this simulation, $k/k_L = 10$ and $\varepsilon_L = 0.1$ as in the previous
two simulations.
The modulations of short-wave wavenumber, amplitude, and steepness in response
to such long-wave group is shown in Fig. \ref{fig:modulation_3panel_groups}.
Like in the case of a linear ramp, the modulation pattern here is locked in to
the long-wave crest and peaks at approximately 1.2 for short-wave steepness of 0.1.
After the peak of the long-wave group passes, the short-wave modulation recedes
back toward the resting state.

\section{Discussion}
\label{section:discussion}

The solutions presented here, analytical and numerical alike, are valid only
for certain scales of short and long waves.
Specifically, the wave action conservation equation, introduced by
\citet{bretherton1968wavetrains}, requires that the ambient velocity is slowly
varying on the time scales of short waves, \textit{i.e.}
$|\partial U / \partial x| \ll k |U|$
and $|\partial U / \partial t| \ll \sigma |U|$.
This implies $k_L \ll k$ and $\sigma_L \ll \sigma$, respectively.
The same requirement is imposed on the properties of short waves, including
wavenumber, amplitude, and effective gravity.
Another requirement is for the short-waves to remain linear, which implies
$a k \ll 1$.
These requirements have been evaluated for the solutions presented here by
quantifying the homogeneity and stationarity of each relevant property of the
flow in \S\ref{sec:wave_action_conservation}.
Furthermore, the short waves examined here do not include the effects of
surface tension (gravity-capillary waves) or intermediate or shallow water,
although the solutions can be extended to those conditions, as
\citet{phillips1981dispersion} did, for example.
Within these limits, the crest-action conservation equations provide an
elegant and simple framework to study the hydrodynamic modulation
of short waves by longer waves.
This approach, of course, isolates this particular process from other
dominant modulation and nonlinear transfer processes, such as wave growth
by wind, wave dissipation, and aerodynamic sheltering by the longer waves,
that are otherwise present in most measurements
\citep[e.g.,][]{plant1986two,laxague2017gravity}.
An alternative approach using a fully-nonlinear velocity potential flow solver
that allows for two-way interaction between short and long waves would provide
additional insight into the modulation of the long waves by the short waves.

Although in this paper we only explored the solutions for $k_L = 1$ and $k=10$
the results are not sensitive to different values of $k_L$, provided the same
long-wave steepness $\varepsilon_L = a_L k_L$.
Numerical simulations with $k/k_L = 10$ and $k/k_L = 100$ show little
sensitivity to the ratio $k/k_L$ (not shown), consistent with the results of
\citet{longuet1987propagation}.
Specifically, the short-wave steepness modulation is 1\% larger at
$\varepsilon_L = 0.27$ for $k/k_L = 10$ compared to $k/k_L = 100$, and 4\%
larger at $\varepsilon_L = 0.4$.
This suggests that as long as $k/k_L \gg 1$, the modulation amplitude is
largely independent of the ratio $k/k_L$.
These results can be reproduced with the companion numerical model.
Note, however, according to \S\ref{sec:wave_action_conservation}, that the
wave action conservation requirements are more easily satisfied for larger
$k/k_L$.
Also not considered here are the short-wave trains that are oblique to the
longer waves, which is relatively straightforward to include
\citep{peureux2021unsteady}.

Despite several studies that provided qualitatively similar steady solutions
for the short-wave modulation by long waves
\citep{longuet1960changes,phillips1981dispersion,longuet1987propagation,henyey1988energy,zhang1990evolution},
it was not obvious from the governing prognostic equations that the steady
solutions should exist generally.
\citet{peureux2021unsteady} examined this problem numerically and found that
the solutions are stable only if the short-wave field is initialised from
a prior steady solution of the form of that from \citet{longuet1960changes}.
If the short-wave field is instead initialised from a uniform field, they
found that the short-wave amplitude, and thus steepness as well, grow
unsteadily and indefinitely.
Here we demonstrate that the unsteady steepening of short
waves only occurs in the case of a sudden appearance of long waves, where
the velocity fields induced by the long waves act in their full capacity
on the short waves that have not yet had time to lock-in to the long-wave
crests.
In such a scenario, sudden forcing on short waves causes the modulation
pattern to detach itself from the long-wave crests and to propagate at the
short-wave group speed.
When this occurs, the modulated short-waves are no longer in a stabilising
resonance with the surface convergence and divergence induced by the long-wave
orbital velocity, and each long-wave passage further destabilises the short-wave
field, causing it to steepen indefinitely.

Another important consideration of the hydrodynamic modulation process is
whether it should be accounted for in phase-averaged spectral wave
models \citep{group1988wam,tolman1991third,donelan2012modeling} used for global
and regional wave forecasting.
Second-order wave spectra have historically been provided by some wave models
due to the importance for understanding radar altimetry data, however only in a
diagnostic capacity \citep[e.g.,][]{janssen2009some}.
The framework by \citet{janssen2009some}, based on the Hamiltonian and
the action series expansion by \citet{zakharov1968stability}, is an alternative
and more general approach to nonlinear wave evolution than the one described
here.
Currently, most spectral wave models used for research or forecasting
do not consider the preferential steepening of short waves in the context
of their dispersion properties or the wind input via the short-wave phase speed.
Some dissipation source functions use the cumulative mean squared slope
to quantify preferential dissipation of short waves due to this effect
\citep{donelan2012modeling,romero2019distribution}.
In the early days of spectral wave modelling, \citet{willebrand1975energy}
examined the effect of a wave spectrum on the short-wave group velocity
and found the modulation effect to be up to several percent.
For a practical implementation, this would incur an additional loop over all
frequencies longer than the considered short-wave frequency.
It was thus not deemed practical to include this effect in wave models at the
time.
However, with the recent advances in computing power, machine learning for
acceleration of complex functions, and phase-resolved modelling of short waves,
considering the effect of wave-modulation beyond the simple cumulative mean
squared slope deserves re-examination.

\section{Conclusions}
\label{section:conclusions}

In this paper we revisited the problem of short-wave modulation by long waves
from a wave action conservation perspective.
Using the linearised wave crest and action conservation equations coupled with
nonlinear solutions for the effective gravity, we derived steady, nonlinear, analytical
solutions for the modulation of short wave wavenumber, amplitude, steepness,
intrinsic frequency, and phase speed.
The latter two were previously examined in the context of resonant interactions
\citep{longuet1962resonant,longuet1962phase}, a modulation process distinct from
the one discussed here.
Both effects, however, may be important for correctly calculating the wind input
into short waves, which is proportional to the wind speed relative to the wave
celerity.
The approximate steady solutions yield higher wavenumber and amplitude
modulation than that of \citet{longuet1960changes}, and somewhat lower than
the numerical solutions of \citet{longuet1987propagation} and
\citet{zhang1990evolution}.
This is mainly due to the analytical solution requiring the crest and action
conservation equations to be linearised, and the nonlinearity arising due
to the evaluation of short waves on the surface of the long wave (as opposed to
the mean water level) and due to nonlinear effective gravity.
In the steady solutions, the short-wave steepness modulation is dominated by
the wavenumber modulation (5/8), followed by the wave action modulation (1/4),
and is least affected by the effective gravity modulation (1/8).
The homogeneity and stationarity criteria for wave action conservation are
evaluated for both the analytical and numerical solutions.
The stationarity is a stronger criterion than homogeneity, and is found to be
strongly satisfied ($> 99\%$ stationary) for very gently sloped waves and weakly
satisfied ($> 90\%$ stationary) for steeper waves.

The numerical solutions of the full wave crest and action conservation equations
are consistent with the analytical solutions for $\varepsilon_L \lesssim 0.2$,
and are similar in magnitude to the numerical solutions of
\citet{longuet1987propagation} and \citet{zhang1990evolution}.
The simulations also reveal that the unsteady growth of the short-wave steepness
reported by \citet{peureux2021unsteady} is due to the sudden appearance of
long waves in a homogeneous short-wave field, a scenario that is unlikely to
be common in the open ocean.
A more likely condition, such as that of long-wave groups, is only mildly
destabilising for the short waves.
These results suggest that for the long-wave steepness $\varepsilon_L \lesssim 0.2$,
the analytical steady solutions presented here are sufficient to describe the
modulation of short waves by long waves.
For moderately steep long waves ($0.2 \lesssim \varepsilon_L \lesssim 0.3$),
numerical solutions of non-linearised crest and action conservation equations
are necessary to properly capture the short-wave steepness modulation.
For steeper long waves ($\varepsilon_L \gtrsim 0.3$), the short-wave properties can
no longer be considered stationary \citep{bretherton1968wavetrains};
the wave action conservation equation is thus not valid in this context.
It may be worthwhile to revisit a more systematic inclusion of the hydrodynamic
modulation effects in phase-averaged spectral wave models, beyond the simple
cumulative mean squared slope for wave dissipation, but also for modulating
the dispersion and wind input into short waves.

\backsection[Acknowledgements]{
I am thankful to Nathan Laxague, Fabrice Ardhuin, Nick Pizzo, and Peisen Tan
for fruitful discussions.
I also thank three anonymous reviewers for their expertise that helped improve
this paper.
}

\backsection[Funding]{
  Milan Curcic was partly supported by the National Science Foundation
  grant AGS-1745384, Transatlantic Research Partnership by the French Embassy
  in the United States and the FACE Foundation, and the Office of Naval Research
  grants N000142012102 (Coastal Land Air-Sea Interaction DRI) and
  N000142412598 (in collaboration with SASCWATCH: Study on Air-Sea Coupling with
  WAves, Turbulence, and Clouds at High winds MURI).
}

\backsection[Declaration of interests]{The author reports no conflict of interest.}

\backsection[Data availability statement]{
  The code to run the numerical simulations and generate the figures in this
  paper is available at \url{https://github.com/wavesgroup/wave-modulation-paper}.
  The numerical model is available as a standalone Python package at
  \url{https://github.com/wavesgroup/2wave}.
  A Python implementation of SSGW by \citet{clamond2018accurate} is available
  at \url{https://github.com/wavesgroup/ssgw}.
}

\backsection[Author ORCIDs]{M. Curcic, https://orcid.org/0000-0002-8822-7749}


\appendix

\section{Derivation of the governing equations}
\label{appendix:derivation}

Here we derive the governing equations for the short waves
(\ref{eq:dispersion})-(\ref{eq:wave_action}) and discuss their asymptotic
consistency.

\subsection{Wave dispersion}

An irrotational flow has a velocity potential such that the continuity can be
expressed as:

\begin{equation}
  \nabla^2 \phi = 0
  \label{eq:laplace_equation}
\end{equation}
Boundary conditions that allow solving for $\phi$ are the dynamic and kinematic
free surface boundary conditions, respectively:

\begin{equation}
  \frac{\partial \phi}{\partial t} + \frac{1}{2} (\nabla{\phi})^2 + gz = 0 \quad \text{at} \quad z = \eta(x,t)
  \label{eq:dynamic_free_surface_boundary_condition}
\end{equation}

\begin{equation}
  \frac{\partial \eta}{\partial t} + \frac{\partial \phi}{\partial x} \frac{\partial \eta}{\partial x} + \frac{\partial \phi}{\partial z} = 0 \quad \text{at} \quad z = \eta(x,t)
  \label{eq:kinematic_free_surface_boundary_condition}
\end{equation}
Also required is the bottom boundary condition:

\begin{equation}
  \frac{\partial \phi}{\partial z} = 0 \quad \text{at} \quad z = -h
\end{equation}
In the \citet{stokes1847} perturbation approach, the velocity potential and
elevation are expanded in a power series of the wave steepness
$\varepsilon = ak$ as:

\begin{equation}
  \phi = \sum_{n=1}^{\infty} \varepsilon^n \phi_n
\end{equation}
\begin{equation}
  \eta = \sum_{n=1}^{\infty} \varepsilon^n \eta_n
\end{equation}
The power series expansion requires $\varepsilon \ll 1$.
The velocity potential and elevation series are truncated at the desired order
and inserted into the dynamic free surface boundary condition (\ref{eq:dynamic_free_surface_boundary_condition}).
Up to the third order, the solutions for the velocity potential and the
corresponding velocity components are (\ref{eq:phi})-(\ref{eq:W_L}), and the
surface elevation is (\ref{eq:eta_stokes}).
Evaluating these quantities to higher orders in $\varepsilon$ is possible at
the expense of higher complexity and with diminishing returns in accuracy.

Combining the solutions for the potential and the surface elevation with the
kinematic free surface boundary condition (\ref{eq:kinematic_free_surface_boundary_condition})
yields the dispersion relation, which in the linear case and in deep water is
(\ref{eq:dispersion}).
At the third order in $\varepsilon$, the frequency begins to depend on the wave
steepness $\varepsilon$ as well:

\begin{equation}
  \omega = \sqrt{g k} \left(1 + \frac{1}{2} \varepsilon^2 + \mathcal{O}(\varepsilon^4)\right) + U k
  \label{eq:dispersion_nonlinear}
\end{equation}
In this paper we constrain the analysis framework to the linear case for short
waves, as this permits the analytical solutions in $\S\ref{section:analytical_solutions}$.
The Doppler shift by the ambient velocity $U$ is unaffected by the nonlinear
effects in the solution.

\subsection{Conservation of wave crests}

A system with a well defined phase function $\psi(x,t)$ also has the wavenumber
and frequency defined as its gradients with respect to space and time, respectively:

\begin{equation}
  k = \frac{\partial \psi}{\partial x}
  \label{eq:wavenumber_from_phase}
\end{equation}

\begin{equation}
  \omega = - \frac{\partial \psi}{\partial t}
  \label{eq:frequency_from_phase}
\end{equation}
implying $\psi = kx - \omega t$.
Differentiating (\ref{eq:wavenumber_from_phase})-(\ref{eq:frequency_from_phase})
with respect to time and space, respectively, and then eliminating $\psi$ yields:

\begin{equation}
  \frac{\partial k}{\partial t} + \frac{\partial \omega}{\partial x} = 0
  \label{eq:wavenumber_conservation}
\end{equation}
which is the crest conservation equation (\ref{eq:wave_crests}).
That this is the wavenumber conservation equation with the group speed as the
transport velocity can be recognized by writing:

\begin{equation}
  \frac{\partial k}{\partial t} + \frac{\partial \omega}{\partial k} \frac{\partial k}{\partial x} = 0
\end{equation}
This conservation equation is valid regardless of the order of nonlinearity
considered, as long as $\psi$ is well defined.
Nonlinearity in terms of $\varepsilon$ can, however, enter (\ref{eq:wavenumber_conservation})
through nonlinear forms of $\omega$ such as (\ref{eq:dispersion_nonlinear}).

\subsection{Conservation of wave action}
\label{appendix:wave_action_conservation}

The conservation of wave action (\ref{eq:wave_action}) is most readily derived
following the variational approach by \citet{whitham1965general}.
Begin with a statement that infinitesimal variations of a Lagrangian of a system
over a spatial domain and time interval must vanish:

\begin{equation}
  \delta \iint L\, dx\,dt = 0
\end{equation}
Here, $L$ is the instantaneous Lagrangian density, a scalar quantity that
corresponds to the difference between the kinetic and potential energy of the
system:

\begin{equation}
  L = \int_{-h}^{\eta} \left[ \frac{1}{2} (\nabla{\phi})^2 + gz \right] dz
  \label{eq:lagrangian_density_energy_difference}
\end{equation}
\citet{luke1967variational} showed that the equations for irrotational
surface waves follow from the variational approach if the Lagrangian $L$ is
defined instead as:

\begin{equation}
  L = \int_{-h}^{\eta} \left[ \frac{\partial \phi}{\partial t} + \frac{1}{2} (\nabla{\phi})^2 + gz \right] dz
  \label{eq:lagrangian_density_luke}
\end{equation}
(\ref{eq:lagrangian_density_energy_difference}) and (\ref{eq:lagrangian_density_luke})
are equivalent in the sense that they both yield (\ref{eq:laplace_equation}),
however, varying the latter is necessary to attain the correct boundary conditions
\citep{whitham1967non}.
Notice that for water waves the instantaneous Lagrangian being zero is
equivalent to the Bernoulli's equation, which is also the dynamic free surface
boundary condition.

Then, Whitham's averaged variation principle is obtained by averaging the
instantaneous Lagrangian $L$ over one wave period:

\begin{equation}
  \mathcal{L} = \frac{1}{2\pi} \int_0^{2\pi} L\ d\psi
  \label{eq:averaged_lagrangian}
\end{equation}
where $\mathcal{L}$ is the averaged Lagrangian density.
The averaged variational principle then states:

\begin{equation}
  \delta \iint \mathcal{L}\, dx\,dt = 0
\end{equation}
The averaged Lagrangian for water waves is obtained by integrating
(\ref{eq:lagrangian_density_luke}) over depth, and then averaging over the phase
(\ref{eq:averaged_lagrangian}).
In the deep water limit, it is:

\begin{equation}
  \mathcal{L} = \frac{1}{4} \left( 1 - \frac{\omega^2}{gk} \right) g a^2 + \mathcal{O}(\varepsilon^4)
\end{equation}
which, after recognizing the wave energy$E = \frac{1}{2} g a^2$, yields:

\begin{equation}
  \mathcal{L} = \frac{1}{2} \left( 1 - \frac{\omega^2}{gk} \right) E + \mathcal{O}(\varepsilon^4)
  \label{eq:averaged_lagrangian_deep_water}
\end{equation}

Whitham's conservation equations for water waves are:

\begin{equation}
  \frac{\partial \mathcal{L}}{\partial a} = 0
\end{equation}

\begin{equation}
  \frac{\partial}{\partial t} \left(\frac{\partial \mathcal{L}}{\partial \omega}\right) - \frac{\partial}{\partial x} \left(\frac{\partial \mathcal{L}}{\partial k}\right) = 0
  \label{eq:whitham_conservation_equation}
\end{equation}
Inserting (\ref{eq:averaged_lagrangian_deep_water}) into (\ref{eq:whitham_conservation_equation})
and defining:

\begin{equation}
  \frac{\partial \mathcal{L}}{\partial \omega} = \frac{E}{\sigma} \equiv N
  \label{eq:adiabatic1}
\end{equation}
yields the conservation of wave action:

\begin{equation}
  \frac{\partial}{\partial t} \left[ N + \mathcal{O}(\varepsilon^4) \right] +
  \frac{\partial}{\partial x} \left[ (C_g + \mathcal{O}(\varepsilon^2)) N + \mathcal{O}(\varepsilon^4) \right] = 0
\end{equation}
Due to the $\mathcal{O}(\varepsilon^2)$ correction to the group speed, the
overall truncation error for the wave action balance is $\mathcal{O}(\varepsilon^2)$
as well:

\begin{equation}
  \frac{\partial N}{\partial t} +
  \frac{\partial (C_g N)}{\partial x} = \mathcal{O}(\varepsilon^2)
  \label{eq:wave_action_conservation_nonlinear}
\end{equation}
which was introduced as a governing equation in \S\ref{section:governing_equations}.
Although the truncation error of $\mathcal{O}(\varepsilon^2)$ arising from $C_g$
seems to have been omitted in the original derivation by \citet{whitham1967non},
it appeared correctly later in \citet{whitham1974linear}.
An important nuance about applying the nonlinear dispersion here is that the
secondary (wave-induced) mean flow and set-up effects can be ignored in the
deep water limit; this is discussed by \citet{whitham1974dispersive}.
The terms proportional to $\varepsilon^4$ do not appear in the averaged Lagrangian
(\ref{eq:averaged_lagrangian_deep_water}) until the third-order Stokes waves
are considered, for which the dispersion gains a $\mathcal{O}(\varepsilon^2)$
correction.
In that case, the conserved quantity is:

\begin{equation}
  \frac{\partial \mathcal{L}}{\partial \omega} = \frac{E}{\sigma} (1 + \frac{1}{2} \varepsilon^2)
  \label{eq:adiabatic2}
\end{equation}
and the group speed gains a nonlinear factor in $\varepsilon^2$ as well.
(\ref{eq:wave_action_conservation_nonlinear}), then, maintains the same form,
but with the truncation error of $\mathcal{O}(\varepsilon^4)$.
This derivation, as well as the averaged Lagrangian principle applied to the
wave-induced mean flow over arbitrary water depths, is detailed at greater length
by \citet{whitham1967non}.

To consider the effect of slowly varying ambient currents,
\citet{bretherton1968wavetrains} carried out the same derivation except that
they propagated the ambient current derivatives $\partial U / \partial t$ and
$\partial U / \partial x$ through the averaged Lagrangian.
To allow these derivatives to drop out of the averaging integral, the authors
make an argument that the variation must be sufficiently small, no larger than
$\mathcal{O}(\frac{1}{\sigma U} \frac{\partial U}{\partial t})$ and
$\mathcal{O}(\frac{1}{k U} \frac{\partial U}{\partial x})$, respectively.
Taking into consideration the asymptotic limits in terms of the wave steepness
$\varepsilon$ as well as the variation of the ambient current, the wave action
balance can be written as:

\begin{equation}
  \frac{\partial N}{\partial t} +
  \frac{\partial}{\partial x} \left[ \left(C_g + U\right) N \right] =
  \mathcal{O}\left(\varepsilon^4, \frac{1}{\sigma U} \frac{\partial U}{\partial t}, \frac{1}{k U} \frac{\partial U}{\partial x}\right)
\end{equation}
\S\ref{sec:wave_action_conservation} deals specifically with quantifying the
ambient velocity variation in the context of long-wave orbital velocities.

There is an alternative derivation of the wave action conservation that was
pointed out by \citet{bretherton1968wavetrains}.
Starting from the wave energy conservation in non-uniform currents derived by
\citet{longuet1961changes}

\begin{equation}
  \frac{\partial E}{\partial t} +
  \frac{\partial}{\partial x} \left[ \left(C_g + U\right) E \right] +
  S_{x} \frac{\partial U}{\partial x} = 0
  \label{eq:energy_conservation_nonuniform_currents}
\end{equation}
where $S_x$ is the wave radiation stress, and combining it with the dispersion
relation (\ref{eq:dispersion}) and conservation of crests
(\ref{eq:wave_crests}), they show that (\ref{eq:energy_conservation_nonuniform_currents})
is equivalent to the wave action conservation (\ref{eq:wave_action}).
This derivation is found in the Appendix of \citet{bretherton1968wavetrains},
and was later recognized by \citet{whitham1974linear} as well.

In summary, asymptotically consistent sets of dispersion and action conservation
equations vary depending on the level of nonlinearity considered.
In the case of linear waves, the dispersion relation (\ref{eq:dispersion}),
the averaged Lagrangian (\ref{eq:averaged_lagrangian_deep_water}), the adiabatic
quantity (\ref{eq:adiabatic1}) and the wave action conservation
(\ref{eq:wave_action_conservation_nonlinear}) are asymptotically consistent.
For third-order Stokes waves, the dispersion relation
(\ref{eq:dispersion_nonlinear}), the adiabatic quantity (\ref{eq:adiabatic2}),
together with the group speed that includes a $\mathcal{O}(\varepsilon^2)$ correction,
are governed by the action conservation law
(\ref{eq:wave_action_conservation_nonlinear}) and are asymptotically consistent.
This is summarized in Table \ref{table:asymptotic_consistency}.

\begin{table}
\begin{center}
\def~{\hphantom{0}}
\begin{tabular}{cccc}
Order & Frequency  & Group speed & Wave action conservation \\
\hline
1, 2 & $\sqrt{gk}(1 + \mathcal{O}(\varepsilon^2))$ & $\frac{1}{2}\sqrt{\frac{g}{k}}(1 + \mathcal{O}(\varepsilon^2))$ & $\frac{\partial N}{\partial t} + \frac{\partial (C_g N)}{\partial x} = \mathcal{O}(\varepsilon^2)$ \\
\hline
3 & $\sqrt{gk}(1 + \frac{1}{2}\varepsilon^2 + \mathcal{O}(\varepsilon^4))$ & $\frac{1}{2}\sqrt{\frac{g}{k}}(1 + \frac{5}{2}\varepsilon^2 + \mathcal{O}(\varepsilon^4))$ & $\frac{\partial N}{\partial t} + \frac{\partial (C_g N)}{\partial x} = \mathcal{O}(\varepsilon^4)$ \\
\hline
\end{tabular}
\caption{
  Asymptotically consistent sets of dispersion and wave action relations for
  deep water and their respective truncation errors in terms of the wave
  steepness $\varepsilon$, for the first three Stokes expansion orders.
}
\label{table:asymptotic_consistency}
\end{center}
\end{table}

\begin{figure}
\centering
\includegraphics[width=0.9\textwidth]{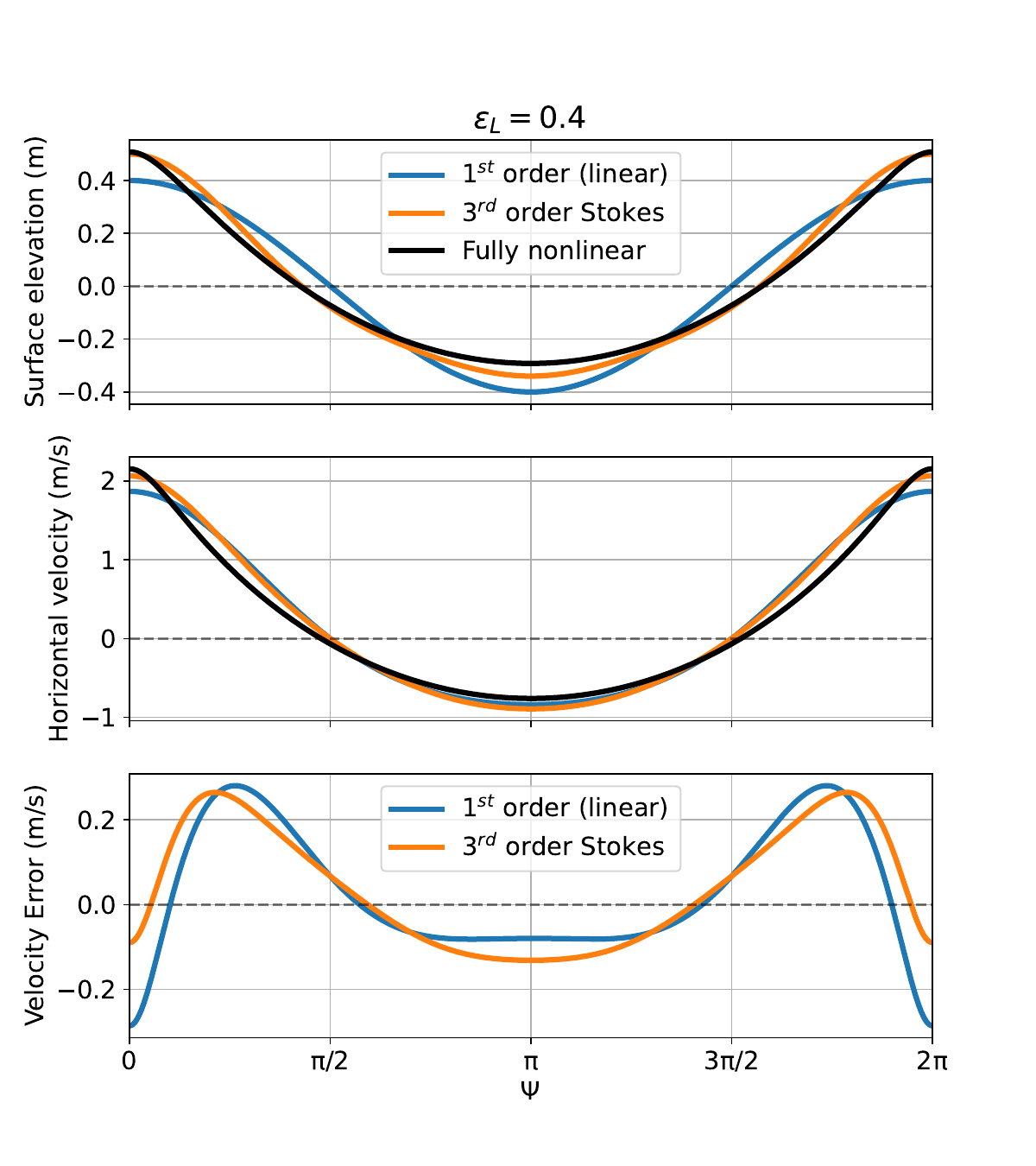}
\caption{
  Surface elevation (top) and horizontal velocity (middle) for the linear wave
  (blue), the third-order Stokes wave (orange), and the fully-nonlinear wave
  (black) with steepness of $\varepsilon_L = 0.4$.
  The bottom panel shows the surface velocity error for the linear wave (blue)
  and the third-order Stokes wave (orange), relative to the fully-nonlinear
  wave solution.
}
\label{fig:velocity_error_by_phase}
\end{figure}

\section{Surface velocity errors}
\label{appendix:surface_velocity_errors}

The long-wave induced velocities (\ref{eq:phi})-(\ref{eq:W_L}) are accurate to
$\mathcal{O}(\varepsilon_L^3)$ when evaluated at the mean water level.
However, in this paper we depart from the mean water level and evaluate the
velocities at the surface of the long wave.
This introduces additional errors at $\mathcal{O}(\varepsilon_L^2)$.
To quantify this error exactly, we compare the velocities evaluated at the
surface for a linear (small-amplitude) wave and a third-order Stokes wave,
with the surface velocity of a fully-nonlinear wave
(Fig. \ref{fig:velocity_error_by_phase}).
The elevation and the surface velocity for the fully-nonlinear wave is
calculated numerically using the method by \citet{clamond2018accurate},
which relies on Petviashvili's iterations and in this case uses 2048 Fourier
modes to represent the wave.
Both the first-order and the third-order waves underestimate the surface velocity
at the crest and in the trough, and overestimate it on the front and rear faces
of the crest.

\begin{figure}
\centering
\includegraphics[width=0.8\textwidth]{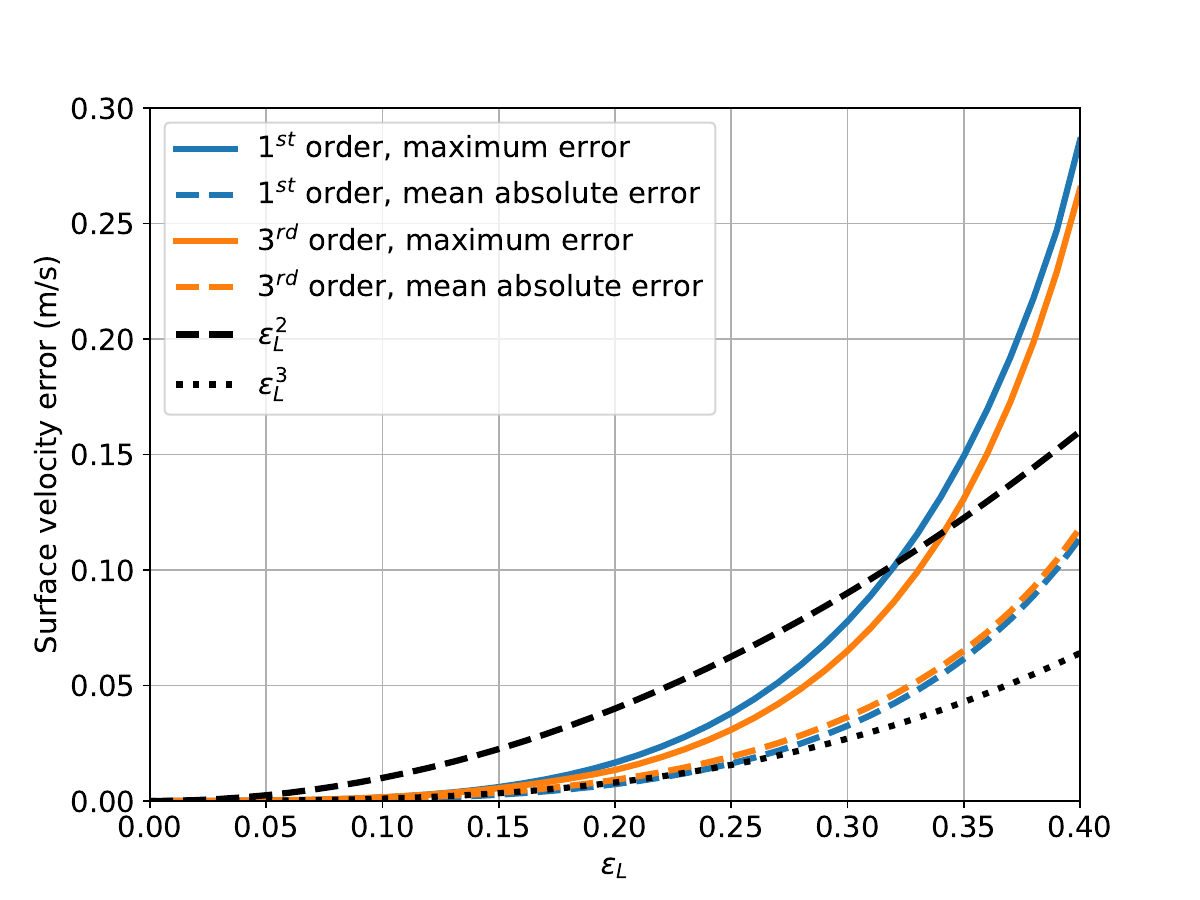}
\caption{
  Surface velocity error as function of wave steepness $\varepsilon_L$,
  for the linear wave (blue) and the 3$^{rd}$ order Stokes wave (orange).
  Maximum and mean errors are indicated with solid and dashed lines,
  respectively.
  For reference, $\varepsilon_L^2$ and $\varepsilon_L^3$ curves are shown in
  black dashed and dotted lines, respectively.
}
\label{fig:velocity_error_by_ak}
\end{figure}

Maximum values of surface velocity errors for both the linear and the
third-order Stokes wave asymptote to $\varepsilon_L^3$ for small
$\varepsilon_L$ (Fig. \ref{fig:velocity_error_by_ak}).
However, they approach and exceed $\varepsilon_L^2$ at $\varepsilon_L \approx 0.3$
and beyond.
Although the maximum error is $\approx 10-20\%$ smaller for the third-order
Stokes wave relative to the linear wave, its average error is slightly larger.
The main takeaway from this figure is that evaluating the surface velocity
of a linear wave carries an error no greater than $\varepsilon_L^2$ for
$\varepsilon_L < 0.32$.
Field measurements of near-surface orbital velocities by
\citet{laxague2020observations} qualitatively support this result, although
they find larger departures from the linear theory in stronger winds, that is,
steeper waves.

\bibliographystyle{jfm}
\bibliography{references}

\end{document}